\documentclass[12pt]{article}
\usepackage[top=0.7in,bottom=1in,left=1.1in,right=1.1in,includehead]{geometry}
\usepackage{epsfig,amsmath,amsfonts,verbatim,cite}
\usepackage{lmodern}
\usepackage[T1]{fontenc}
\usepackage{mathtext,marvosym,textcomp}
\usepackage{epsfig,amsmath,amsfonts,amssymb}
\usepackage{mathtools}
\usepackage{slashed}
\usepackage[usenames,dvipsnames,svgnames,table]{xcolor}
\usepackage{tikz}
\usetikzlibrary{arrows}
\usetikzlibrary{shapes.misc}
\usetikzlibrary{shapes,positioning,matrix,arrows}
\tikzstyle{every picture}+=[remember picture]
\tikzstyle{na} = [baseline=-.5ex]
\tikzstyle{format} = [rectangle,
                      thick,
                      minimum size=2cm,
                      draw=red!00!black!50,
                      top color=white,
                      bottom color=red!00!black!00,
                      text centered,
                      font=\fontsize{10}{10}\selectfont,
                      on grid]
\tikzstyle{format1} = [rectangle,
                      thick,
                      dashed,
                      minimum size=2cm,
                      draw=red!00!black!50,
                      top color=white,
                      bottom color=red!00!black!00,                     
                      text centered,
                      font=\fontsize{10}{10}\selectfont,
                      on grid]
\tikzstyle{format2} = [font=\fontsize{10}{10}\selectfont,
                      on grid]
\tikzset{cross/.style={cross out, draw=black, minimum size=2*(#1-\pgflinewidth), inner sep=0pt, outer sep=0pt},
cross/.default={5pt}}
 \usepackage[ normalem]{ulem}
 \usepackage{tocloft}
 
 \usepackage[debug,pageanchor=false]{hyperref}
\hypersetup{colorlinks=true,linktocpage,breaklinks,
            urlcolor=blue,
            linkcolor=red,
            citecolor=blue
            }

\numberwithin{equation}{section}

\def\a{\alpha} \def\b{\beta} \def\g{\gamma} \def\d{\delta} \def\e{\epsilon}
  \def\h{\eta} 
    \def\m{\mu}
\def\n{\nu}    \def\r{\rho}
 \def\s{\sigma}

   \def\L{\Lambda} 
  \def\P{\Pi}

\def\ba{\bar{a}}\def\bb{{\bar{b}}}\def\bc{\bar{c}}\def\bd{\bar{d}}

\def\fr{\frac}  \def\dt{\partial}

\def\hh{\hat{h}}
\def\ph{\phantom}

\def\mF{\mathcal{F}}

\def\mZ{\mathcal{Z}}
\def\mG{\mathcal{G}}

\def\mH{\mathcal{H}}
\def\mL{\mathcal{L}}

\def\tx{\tilde{x}}

\def\tg{\tilde{g}}
\def\tk{\tilde{k}}

\def\vac{|0\rangle}
\def\tm{\times}

\newcommand\bqa {\begin{eqnarray}}
\newcommand\eqa {\end{eqnarray}}

\newcommand{\bear}{\begin{array}}
\newcommand{\enar}{\end{array}}

\newcommand{\be}{\begin{equation}}
\newcommand{\ee}{\end{equation}}
\def\bea{\begin{eqnarray}}
\def\eea{\end{eqnarray}}

\newcommand{\cC}{\mathcal{C}}			

\newcommand{\gM}{\mathcal{H}}

\newcommand{\hmu}{{\hat{\mu}}}
\newcommand{\hnu}{{\hat{\nu}}}

\newcommand{\aw}{\alpha}
\newcommand{\bw}{\beta}

\newcommand{\gV}{\mathcal{A}}

\newcommand{\lambdab}{\lambda_{\text{brane}}}
\newcommand{\blambdab}{\bar{\lambda}_{\text{brane}}}
\newcommand{\cc}{\tilde{c}}

\renewcommand{\hh}{h}

\renewcommand{\hh}{h}

\begin{document}
\renewcommand{\contentsname}{}
\renewcommand{\refname}{\begin{center}References\end{center}}
\renewcommand{\abstractname}{\begin{center}\footnotesize{\bf Abstract}\end{center}} 
 \renewcommand{\cftdot}{}

\begin{titlepage}
\ph{preprint}

\vfill

\begin{center}
   \baselineskip=16pt
   {\large \bf 
   Five-brane actions in double field theory    
   }
   \vskip 2cm
    Chris D. A. Blair$^\star$\footnote{\tt cblair@vub.ac.be }
    Edvard T. Musaev$^\dagger{}^\bullet$\footnote{\tt edvard.musaev@aei.mpg.de}
       \vskip .6cm
             \begin{small}
                          {\it
                          $^\star$Theoretische Natuurkunde, Vrije Universiteit Brussel, and the International Solvay Institutes, \\
                          Pleinlaan 2, B-1050 Brussels, Belgium\\[0.5cm]
                          $^\dagger$Max-Planck-Institut f\"ur Gravitationsphysik (Albert-Einstein-Institut)\\
                          Am M\"uhlenberg 1, DE-14476 Potsdam, Germany\\[0.5cm] 
                           $^\bullet$Kazan Federal University, Institute of Physics\\
                          Kremlevskaya 16a, 420111, Kazan, Russia                         
                          } \\ 
\end{small}
\end{center}

\vfill 
\begin{center} 
\textbf{Abstract}
\end{center} 
\begin{quote}
We construct an action for NSNS 5-branes which is manifestly covariant under ${\rm O}(d,d)$. 
This is done by doubling $d$ of the spacetime coordinates which appear in the worldvolume action.
By formulating the DBI part of the action in a manner similar to a ``gauged sigma model'', only half the doubled coordinates genuinely appear. 
Our approach allows one to describe the full T-duality orbit of the IIB NS5 brane, the IIA KKM and their exotic relations in one formalism.
Furthermore, by using ideas from double field theory, our action can be said to describe various aspects of non-geometric five-branes. 
\end{quote} 
\vfill
\setcounter{footnote}{0}
\end{titlepage}

\tableofcontents

\setcounter{page}{2}

\section{Introduction}

It is well known that despite the name string theory describes not only fundamental strings but also D- and NS-branes. These appear as extended objects with tension proportional to $g_s^{-1}$ and $g_s^{-2}$, respectively, while the tension of the fundamental string itself does not scale with $g_s$. Thus these branes are non-perturbative, or solitonic, in nature. 

That we know as much as we do about the non-perturbative objects of string theory and M-theory is in large part thanks to the existence of dualities \cite{Hull:1994ys}. 
This paper arises as part of a broader effort to understand the consequences of acting with T-duality transformations on solitonic NS branes. 
While the action of T-duality on a D$p$-brane simply generates either a D$(p+1)$- or D$(p-1)$-brane, depending on whether the duality is carried out in a direction transverse to the brane or not, 
the action on NS backgrounds produces results which are rather more interesting.
Working at the level of the corresponding supergravity solutions, the T-dual of the NS5 brane along a transverse isometry produces the Kaluza-Klein monopole (or KKM). 
A second T-duality produces a brane -- known as the $5_2^2$ -- which is ordinarily said to be ``exotic'': it is globally defined only up to a non-trivial T-duality transformation and so can be viewed as a T-fold \cite{Hull:2004in}. It is a ``globally non-geometric background''.

It has been argued in \cite{deBoer:2012ma} that such backgrounds are not really ``exotic'' in the full string theory, but are in fact a ubiquitous feature. Their existence can be viewed as being required by U-duality \cite{Obers:1998fb}, and they provide sources for non-geometric Q- and R-fluxes, T-dual to the usual three-form and geometric fluxes sourced by the NS5 and KKM, of interest for compactifications \cite{
Kachru:2002sk,
Shelton:2005cf}. 

The T-duality orbit beginning with the NS5 brane can be written in the notation of \cite{deBoer:2012ma} as
\begin{equation}
\label{orbit}
\begin{aligned}
& 5_2^0 && \longleftrightarrow && 5_2^1 && \longleftrightarrow && 5_2^2 && \longleftrightarrow && 5_2^3,
\end{aligned}
\end{equation}
where the subscript stands for the power of $g_s^{-1}$ in the tension and the superscript counts the number of special circles.
Here $5_2^0$ is just the usual NS5 brane, while $5_2^1$ is the KK monopole, which has three transverse directions and a special isometry direction, as indicated by the superscript. 
Note that in order to carry out T-dualities in the transverse directions to these branes, one needs transverse isometries and so this necessitates smearing the solution at each step.
This smearing changes the corresponding harmonic function. For the exotic $5_2^2$-brane, one ends up with the harmonic function depending logarithmically on the transverse radius and hence not vanishing at infinity. The $5_2^3$-brane is even more non-geometric: one point of view is that this brane is the result of carrying out a T-duality on a direction which is not an isometry, leading to a background which depends on a dual coordinate (``locally non-geometric''). 

Some other problems arise when considering throat behaviour already at the level of KK monopole as described in
\cite{Gregory:1997te}. 
One question concerns the fact that the T-dual of the KK monopole in supergravity is naively a smeared version of the NS5 brane, while in principle one expects the latter to be localised in its transverse directions.
As shown in \cite{Tong:2002rq}, worldsheet instanton corrections in the smeared NS5 background have the effect of localising the NS5. 
Later on, it was realised that one could calculate the worldsheet instanton corrections to the KK monopole background itself, which turn out to imply that the KK monopole is localised not in the special isometry direction but in a dual direction \cite{ Harvey:2005ab, Jensen:2011jna}. 
Similar calculations have been done for the non-geometric $5_2^2$ brane \cite{Kimura:2013fda,Kimura:2013zva, Kimura:2013khz}. So again, explicit dependence on \emph{dual} coordinates appears in the solution, but this time this is suggested by a \emph{worldsheet} calculation.\footnote{For further recent study of the expected localisations, see \cite{Lust:2017jox}.}

Dealing with non-geometry, whether manifesting as non-trivial T-duality monodromy or as dependence on winding coordinates, requires a formalism that goes beyond our usual supergravity framework.
So far the most appropriate formalism to attempt to make sense of dual coordinates has been that of Double Field Theory (DFT), which starts with doubling (a subset of) the spacetime coordinates and considering a theory invariant under the group ${\rm O}(d,d)$ defined on a space parametrized by $Y^M=(Y^i,\tilde Y_i)$ with $M=1,\dots,2d$. All fields of supergravity then can be arranged into various irreducible representations of ${\rm O}(d,d)$, e.g. the NS-NS sector containing the dilaton $\phi$, metric $g$ and the B-field $B$ is described by the so-called generalised metric 
\begin{equation}
\mH_{MN}\in \fr{{\rm O}(d,d)}{ {\rm O}(d)\times {\rm O}(d)},
\end{equation}
usually parametrised by
\be
\gM_{MN} = \begin{pmatrix} g - B g^{-1} B & Bg^{-1} \\ - g^{-1} B & g^{-1} \end{pmatrix} \,,
\label{genmet}
\ee
and by the generalised dilaton
\be
e^{-2d} =e^{-2\phi} \sqrt{|g|} \,.
\label{gendil}
\ee
The dynamics of the theory in the NS-NS sector are provided by an action \cite{Siegel:1993th, Siegel:1993xq, Hull:2009mi, Hull:2009zb, Hohm:2010jy, Hohm:2010pp} 
\begin{equation}
\begin{split} 
S_{DFT}  =&\ \int d^{2d} Y e^{-2d} \Big( 
 4 \gM^{MN}\partial_{M}\partial_{N}d
-\partial_{M}\partial_{N}\gM^{MN} 
-4\gM^{MN}\partial_{M}d\,\partial_{N}d\\
&+ 4 \partial_M \gM^{MN}  \partial_Nd+ \frac{1}{8}\gM^{MN}\partial_{M}\gM^{KL}
\partial_{N}\gM_{KL}-\frac{1}{2}\gM^{MN}\partial_{M}\gM^{KL}
\partial_{K}\gM_{NL}
 \Big) \,,
\end{split}
\end{equation}
which is fixed by invariance under local transformations which provide the notion of ``generalised diffeomorphisms''. Infinitesimally, these combine conventional diffeomorphisms and gauge transformations into a generalised Lie derivative $\L$
and act on an arbitrary (generalised) vector as
\begin{equation}
\d_\L V^M=\L_\L V^M=\L^N\dt_N V^M - V^N \dt_N \L^M + \eta^{MN} \eta_{PQ} \dt_N \L^P V^Q \,,
\end{equation}
where
\be
\eta_{MN} = 
\begin{pmatrix} 
0 & {\bf 1}_{d\times d} \\ 
{\bf 1}_{d\times d} & 0 
\end{pmatrix} 
\label{eta}
\ee
is preserved by ${\rm O}(d,d)$ transformations. Generalised diffeomorphisms themselves can be viewed as infinitesimal local ${\rm O}(d,d)$ transformations (plus a transport term) in the same way that usual diffeomorphisms are associated to the group $\mathrm{GL}(d)$. 
The algebra of such transformations is closed only upon imposing a special condition which restricts dependence of all fields of the theory. This is usually called the section condition, and is given by:
\begin{equation}
\h^{MN}\dt_M \bullet \dt_N \bullet=0.
\end{equation}
with the bullets standing for any expression in fields. This condition has to be imposed by hand to keep the theory consistent and to return back to the normal number of coordinates. 

The remaining bosonic fields of type II supergravity are the RR fields, and these form spinors of ${\rm O}(d,d)$, with the type II theories distinguished by chirality. 
Extensions of DFT to treat the RR fields and then supersymmetry were provided in \cite{Hohm:2011dv, Hohm:2011nu, Jeon:2011vx, Jeon:2011sq, Jeon:2012kd, Jeon:2012hp}.
A ``split'' or ``Kaluza-Klein'' formulation, in which not all directions are doubled, was provided for the NSNS sector in \cite{Hohm:2013nja}. 
Reviews covering these and many other aspects are \cite{Aldazabal:2013sca, Berman:2013eva, Hohm:2013bwa}. 

Since DFT automatically becomes a T-duality covariant theory in the presence of isometries the whole orbit \eqref{orbit} is represented by a single solution of its equations of motion, which is the so-called DFT monopole constructed in \cite{Berman:2014jsa}. 
As dual coordinates are present in the theory, one can further imagine carrying out ``duality transformations'' along directions which are not isometries, leading to configurations which do not violate the section condition (in that no fields depend on both a coordinate and its dual simultaneously) but which involve explicit dependence on dual coordinates which are not part of the spacetime. 

The DFT monopole solution is characterised by a harmonic function $H=H(y_1,y_2,y_3)$, which depends on three coordinates, which can be identified with either geometric or dual coordinates. As was shown in \cite{Bakhmatov:2016kfn}, depending on the way this is done, one ends up with not only the NS5-brane and KK-monopole but also (a generalisation of) the $5_2^2$-, $5_2^3$- and even $5_2^4$-branes. The latter is a co-dimension-0 object and usually is not considered in the analysis of exotic branes. The main results were: i) the harmonic function is well behaved as $y_1^2+y_2^2+y_3^2$ goes to infinity; ii) the backgrounds depend on non-geometric coordinates in precisely the same way as expected from string world-sheet instanton corrections.
Similar results have been found for brane backgrounds of M-theory in \cite{Bakhmatov:2017les} where the exceptional field theory \cite{Berman:2010is, Hohm:2013pua} realising the U-duality group $\mathrm{SL}(5)$ was considered.

So far we have mainly discussed backgrounds of field theories rather than proper dynamical objects. Ideally, we want to also have an effective worldvolume action for a brane, completing the action to
\begin{equation}
S_{full}=S_{fields}+S_{brane}
\end{equation}
such that the full equations of motion produce the correct background as the solution. In other words, the worldvolume effective action acts as a source of the corresponding brane background. Such actions for the NS5 brane and the KK-monopole are well known \cite{
Callan:1991ky, 
Hull:1997kt, 
Bergshoeff:1997gy, 
Bergshoeff:1997ak, 
Bergshoeff:1998ef, 
Eyras:1998hn 
 }.
We note that the IIA NS5 brane involves a self-dual three-form, and so it is more difficult to obtain a genuine action -- a PST form is provided in \cite{Bandos:2000az}.
Effective actions for exotic branes were considered some time ago in \cite{Eyras:1999at} and more recently in \cite{Chatzistavrakidis:2013jqa, Kimura:2014upa, Kimura:2016anf}, based on dualisation of the known effective actions of NS5-brane and KK monopole along isometry directions. 

The aim of the present paper is to construct an effective action describing the full T-duality orbit \eqref{orbit}, which is ${\rm O}(d,d)$ covariant, reproduces the known effective actions upon solving the section condition appropriately and gives the full DFT-monopole background when considered as a source for the DFT action\footnote{Note that the electric counterpart of this solution, the DFT wave \cite{Berkeley:2014nza}, has been shown in \cite{Blair:2016xnn} to be a solution of the combined action $S_{full} = S_{DFT} + S_{DWS}$, where $S_{DWS}$ is an action for a doubled string such as \cite{Tseytlin:1990nb,Tseytlin:1990va,Hull:2004in,Hull:2006va,Blair:2013noa}.}
\begin{equation}
S_{full}=S_{DFT}+S_{eff}.
\end{equation}
We restrict consideration here only to the case of the T-duality orbit starting with the Type IIB NS5-brane as in this case one is able to write the full non-linear action. For the Type IIA case one would have to work in the PST formalism, or restrict the action to its quadratic form.

We will provide a universal DBI term, valid for any number of doubled directions. The form of this DBI term will be a generalisation to doubled space of a form of the KK monopole action \cite{Eyras:1998hn}: this goes by the name of a ``gauged sigma model'', the idea being that one or more of the target space directions is an isometry and the resulting worldvolume scalar (spacetime coordinate) does not appear in the action, effectively by gauging the isometry. Half of our doubled directions will be viewed in this manner. 

We will also discuss the structure of the Wess-Zumino term for the doubled five-brane. 
The NSNS contribution is complicated as it must describe the T-duals of the electromagnetic dual $B_6$ of the B-field, which means the electromagnetic dual of the Kaluza-Klein vector, and other more exotic objects, for which non-linear definitions are not known. However, a linearised description in DFT has been achieved in \cite{Bergshoeff:2016ncb} while the representation theory structure for this part of the WZ term in ${\rm O}(d,d)$ is known thanks to \cite{Bergshoeff:2011zk, Bergshoeff:2011mh, Bergshoeff:2011ee, Bergshoeff:2011se, Bergshoeff:2012ex}. Furthermore, the paper \cite{Bergshoeff:2011zk} constructs a general formula for T-duality covariant WZ terms of solitonic 5-branes for general $d < 10$ by introducing worldvolume field strengths for every ${\rm O}(d,d)$ covariant multiplet of gauge fields.
These results apply for actual reductions, while in DFT we want to keep the dependence on the doubled internal coordinates (subject to the section condition) and so have more complicated gauge transformations to consider. 
Still, the results of \cite{Bergshoeff:2011zk} provide a useful guide to the structures that we expect to appear. 
We will thus provide worldvolume field strengths, which are actually invariant under the gauge symmetries of the DFT RR fields only on contraction with an auxiliary worldvolume ${\rm O}(d,d)$ spinor which we will explain below. This then allows us to write down the RR contribution to the full ${\rm O}(10,10)$ covariant WZ term. We will then discuss how the coupling to the NSNS dual $O(d,d)$ covariant potentials should be realised in our formalism, for $d=2,4$ and $10$ as examples.

This paper is structured as follows.
We begin by introducing the five-brane actions we are interested in. We do this in Section \ref{revs}: writing down the DBI actions for the IIB NS5 brane and its T-duals, the IIA KKM and the IIB $5_2^2$ brane. Here we will also introduce the basic ideas of the doubled formalism that we will use.
Then, in Section \ref{ODDDBI} we write down our ${\rm O}(d,d)$-covariant DBI action, which resembles that of the ``gauged sigma model'' form of the KKM action, and demonstrate its equivalence to the usual actions. 

In Section \ref{Section_WZ} we consider the Wess-Zumino terms of these five-brane actions and comment on the Bianchi identities in which these five-branes appear as sources. Finally, in Section \ref{discussion} we provide a discussion of various aspects of our construction and of possible future work building on the results of this paper.

\section{Review of 5-brane actions, duality and doubling}
\label{revs}

This section serves to introduce the branes we will study in this paper, and the basics of the doubled formalisms we will be using to reformulate the brane actions.

\subsection{5-brane actions}
\label{5branerev}

We begin by reviewing the known actions for the IIB NS5 brane and its T-duals, following the results of \cite{Eyras:1998hn, Chatzistavrakidis:2013jqa}.

The NS5 brane of type IIB theory has a six-dimensional worldvolume action $S = S_{DBI} + S_{WZ}$, consisting of a DBI part and a Wess-Zumino term. 
The latter contains the coupling to the six-form electromagnetic dual of the B-field, and so takes the form
$S_{WZ} = \mu_{NS5} \int d^6 \sigma B_6 + \dots$, where the dots indicate additional couplings to the RR fields such that $S_{WZ}$ is gauge invariant, as we will discuss in Section \ref{Section_WZ}.
For now we will concentrate on the DBI part:
\be
S_{DBI} = T_{5} \int d^6 \sigma e^{-2\phi} \sqrt{ 1 + C_0^2 e^{2\phi} }
\sqrt{ 
- \det \left( \hat g_{\alpha \beta}  - \frac{ e^\phi}{\sqrt{ 1 + C_0^2 e^{2\phi}}} \mG_{\alpha \beta} \right)
} 
\,.
\label{NS5DBI} 
\ee
Here $\phi$ is the dilaton and $C_0$ the RR 0-form. 
The overall $e^{-2\phi}$ term reveals that the NS5 physical tension is $g_s^{-2} T_5$, with the expected string coupling dependence. The worldvolume fields that appear include the scalars $X^{\hmu} ( \sigma)$, corresponding to the usual spacetime coordinates, and a one-form $c_\alpha$. These appear in the pullback of the metric, $\hat g_{\hmu \hnu}$,
\be
\hat g_{\alpha \beta} = \partial_\alpha X^{\hmu} \partial_\beta X^{\hnu} \hat g_{\hmu \hnu} \,,
\ee
and in the gauge invariant pullback of the RR two-form, $\hat C_{\hmu\hnu}$,
\be
\mG_{\alpha \beta} = 2 \partial_{[\alpha} c_{\beta]} + \partial_\alpha X^{\hmu} \partial_\beta X^{\hnu} \hat C_{\hmu \hnu} \,.
\ee
We can obtain an action for the IIA Kaluza-Klein monopole by T-dualising. To describe this, and to provide the connection to the double field theory approach, let us discuss this in some generality. 

We will be interested in either T-dualising or doubling $d$ directions. 
Let us group the 10-dimensional coordinates as $X^{\hmu} = (X^\mu, Y^i)$, with $i$ the $d$-dimensional coordinate index and $\mu$ the $D=(10-d)$-dimensional one.
The following Kaluza-Klein type decomposition is used for the NSNS sector fields:\footnote{When we double all coordinate directions ($d=10$) we will simply write the 10-d metric and B-field as $g_{ij}$ and $B_{ij}$, dropping the hats.}
\be
\begin{array}{cll}
\hat g_{\mu\nu} & =& g_{\mu\nu} + g_{ij} A_\mu{}^i A_\nu{}^j \,,\\
\hat g_{\mu i} & = & g_{ij} A_\mu{}^j \,,\\
\hat g_{ij} & = & g_{ij} \,,
\end{array} 
\quad\quad
\begin{array}{cll} 
\hat B_{\mu\nu} & = & B_{\mu\nu} - A_{[\mu}{}^j A_{\nu]j} + B_{ij} A_\mu{}^i A_\nu{}^j\,, \\
 \hat B_{\mu i} & = & A_{\mu i} + A_\mu{}^j B_{ji}\,, \\
\hat B_{ij} & = & B_{ij}\,.
\end{array} 
\label{SUGRAdecomp}
\ee
Now we can recombine the field components into ${\rm O}(d,d)$ multiplets: a one-form transforming as an ${\rm O}(d,d)$ vector, and the generalised metric which is an element of the coset ${\rm O}(d,d) / {\rm O}(d) \times {\rm O}(d)$:
\be
A_\mu{}^M = \begin{pmatrix}  A_\mu{}^i \\ A_{\mu i} \end{pmatrix} \quad,\quad
\gM_{MN} = \begin{pmatrix} 
g_{ij} - B_{ik} g^{kl} B_{lj} & B_{ik} g^{kj} \\
- g^{ik} B_{kj} & g^{ij} 
\end{pmatrix} \,,
\label{AHsplit} 
\ee
as well as the following scalars under ${\rm O}(d,d)$:
\be
g_{\mu\nu} \quad,\quad B_{\mu\nu} \quad,\quad e^{-2d} = e^{-2\phi} \sqrt{|g|}\,,
\label{gbdsplit} 
\ee
where here $|g|  \equiv \det g_{ij}$. 
General ${\rm O}(d,d)$ transformations are those which preserve the ${\rm O}(d,d)$ structure $\eta$, taken to be as in \eqref{eta}.
The standard worldsheet T-duality is a Buscher duality. Such a duality in the direction $x$ acts as permutation interchanging the $V^x$ and $V_{\tilde x}$ components of an ${\rm O}(d,d)$ vector $V^M$. 
Meanwhile, an RR $p$-form transforms into the $(p\pm 1)$-forms of the dual theory, as detailed in Appendix \ref{RRT}.

Using the above multiplets, it is straightforward to T-dualise the NS5 brane action and obtain that of the Kaluza-Klein monopole in type IIA. First, let us write the general decomposition of \eqref{NS5DBI}. We have
\begin{equation}
\label{NS5DBIsplit} 
\begin{aligned}
S_{DBI} =&\ T_{5} \int d^6 \sigma e^{-2\phi} \sqrt{ 1 + C_0^2 e^{2\phi} }\times\\
&\times
\sqrt{ 
- \det \left( g_{\mu\nu} \partial_\alpha X^\mu \partial_\beta X^\nu   
 + g_{ij} D_\alpha Y^i D_\beta Y^j 
 - \frac{ e^\phi \mG_{\alpha \beta}}{\sqrt{ 1 + C_0^2 e^{2\phi}}}  \right)
} 
\,,
\end{aligned}
\end{equation}
where 
\be
D_\alpha Y^i \equiv \partial_\alpha Y^i + \partial_\alpha X^\mu A_\mu{}^i \,,
\ee
and
\be
\begin{split} 
\mG_{\alpha \beta} & = 
2 \partial_{[\alpha} c_{\beta]} 
+ \left( 
\hat C_{\mu\nu}  - 2 \hat C_{[\mu | i |} A_{\nu]}{}^i + \hat C_{ij} A_\mu{}^i A_\nu{}^j 
\right) \partial_\alpha X^\mu \partial_\beta X^\nu 
\\ & \qquad\qquad
+ 2 ( \hat C_{\mu i} - A_\mu{}^j \hat C_{ji} ) \partial_{[\alpha} X^\mu D_{\beta]} Y^i 
+ \hat C_{ij} D_\alpha Y^i D_\beta Y^j \,.
\end{split} 
\ee
Supposing that $d=1$ (and explicitly letting $i=1$) the T-dual expression follows simply from the Buscher rules. 
For instance, we have for the NSNS fields
\be
\tilde g_{11} = \frac{1}{g_{11} }
\quad,\quad
e^{-2\tilde \phi} \sqrt{ \tilde g_{11}} = e^{-2\phi} \sqrt{g_{11}}
\quad,\quad
\tilde A_\mu{}^1 = A_{\mu 1} \quad,\quad
\tilde A_{\mu 1} = A_\mu{}^1 
\ee
where the tilded fields are the T-duals, and in our RR conventions, the relevant T-duality rules are (see Appendix \ref{RRT})
\be
\hat C_{\mu \nu 1} = \hat C_{\mu\nu} + 2 A_{[\mu}{}^1 \hat C_{\nu ]1} 
\quad,\quad
\hat C_{\mu} = \hat C_{\mu 1} + 2 \hat B_{\mu 1} C_0
\quad,\quad
\hat C_1 = C_0 \,.
\ee
We use these to first express \eqref{NS5DBIsplit} in terms of the duals, then dropping the tildes from the fields, we can write the DBI part of the IIA KKM action:
\be
\begin{split} 
S_{DBI} & = T_{5} \int e^{-2 \phi} g_{11} 
\sqrt { 1 + e^{2\phi} \frac{1}{g_{11}} ( \hat C_1 )^2 }
\\ & \qquad\qquad\times
\sqrt{ 
- \det \left(
g_{\mu \nu} \partial_\alpha X^\mu \partial_\beta X^\nu 
+ \frac{1}{\phi_{11}} D_\alpha \tilde Y_1 D_\beta \tilde Y_1
-  \frac{1}{\sqrt{g_{11}}} \frac{ e^\phi  \mG_{\alpha \beta}  }{\sqrt { 1 + e^{2\phi} \frac{1}{g_{11}} ( C_1 )^2 }}\right)
}  
\end{split}
\label{DBIKKM}
\ee
Here $\tilde Y_1$ is the original $Y^1$ appearing in \eqref{NS5DBIsplit}, and we have in the KKM frame that 
\be
D_\alpha \tilde Y_1  \equiv \partial_\alpha \tilde Y_1 + \partial_\alpha X^\mu A_{\mu 1} 
\ee
where $A_{\mu 1}$ are components of the IIA B-field in the decomposition \eqref{SUGRAdecomp}. We also have:
\be
\mG_{\alpha \beta} = 
2 \partial_{[\alpha} c_{\beta]} 
+ \hat C_{\mu\nu 1} \partial_\alpha X^\mu \partial_\beta X^\nu 
- 2 D_{[\alpha} \tilde Y_1 \partial_{\beta]} X^\nu ( \hat C_\nu - A_\nu{}^1 \hat C_1 ) \,,
\ee
were $\hat C_1$ is the $i=1$ component of the RR 1-form.
 
One might wonder how this can be (part of) the action for the KKM, given that it does not seem at all spacetime covariant? The mere fact that we have obtained it by T-duality means that the $i=1$ direction has to be an isometry. (Later on, we will discuss the circumstances in double field theory in which one may not rely on the existence of such isometries.) Indeed, the KKM differs from the more usual fundamental, D- and NS5 branes in that is characterised not simply by its worldvolume and transverse directions, but also has a transverse special isometry direction. Here, this corresponds to the $i=1$ direction. The fact that this is an isometry manifests itself in the absence of the coordinate $Y^1$ from the action \eqref{DBIKKM}. Instead there is a worldvolume scalar, $\tilde Y_1$, which as far as the KKM is concerned can be interpreted as a dual coordinate (reflecting the fact that it is ``originally'' a coordinate in the IIB NS5 action).

A covariant form of the Kaluza-Klein monopole action can be obtained by introducing a Killing vector, $\hat k^{\hmu}$, corresponding to the special isometry direction. In adapted coordinates, $\hat k = \frac{\partial}{\partial Y^1}$, and $|\hat k|^2 \equiv \hat g_{\hmu \hnu} k^{\hmu \hnu} = g_{11}$. 
Let 
\be
\hat \partial_\alpha X^{\hmu} = \partial_\alpha X^{\hmu} - \frac{1}{|\hat k|^2} \hat k^{\hmu} \hat k_{\hnu} \partial_\alpha X^{\hnu} \,,
\ee
such that in adapted coordinates we have $\hat \partial_\alpha X^\mu = \partial_\alpha X^\mu$ and $\hat \partial_\alpha Y^1 = - \frac{\hat g_{\mu 1}}{\hat g_{11}}   \partial_\alpha X^\mu$. 
This ensures that $Y^1$ does not appear in the action. It is easy to check that the following action:
\be
\begin{split} 
S_{DBI} & = T_{5} \int e^{-2 \phi} |\hat k|^2
\sqrt { 1 + e^{2\phi} \frac{1}{|\hat k|^2} (i_{\hat k} \hat C )^2 }
\\ & \qquad\qquad\times
\sqrt{ 
- \det \left(
\hat g_{\hmu \hnu} \hat \partial_\alpha X^{\hmu} \hat \partial_\beta X^{\hnu} 
+ \frac{1}{|\hat k|^2} D_\alpha \tilde Y D_\beta \tilde Y
-  \frac{1}{|\hat k|} \frac{ e^\phi  \mG_{\alpha \beta}  }{\sqrt { 1 + e^{2\phi} \frac{1}{|\hat k|^2} (i_{\hat k} \hat C )^2 }}\right)
}  
\end{split}
\label{DBIKKMcov}
\ee
is the covariantisation of \eqref{DBIKKMcov}, with
\be
D_\alpha \tilde Y \equiv \partial_\alpha \tilde Y - ( i_{\hat k} \hat B )_{\hmu} \partial_\alpha X^{\hmu} 
\ee
and
\be
\mG_{\alpha \beta} = 
2 \partial_{[\alpha} c_{\beta]} 
+ ( i_{\hat k} \hat C) _{\hmu \hnu} \partial_\alpha X^{\hmu} \partial_\beta X^{\hnu} 
- 2 D_{[\alpha} \tilde Y \hat \partial_{\beta]} X^{\hmu} \hat C_{\hmu} \,,
\ee
where $i_{\hat k} T_{\hmu_1 \dots \hmu_p} \equiv \hat k^\mu T_{\mu \mu_1 \dots \mu_p}$.

This form of the action, which if only the metric parts were present corresponds to a gauged sigma model, will be our starting point in constructing manifestly ${\rm O}(d,d)$ covariant 5-brane actions. 

One can further obtain the action for the exotic $5_2^2$ brane by T-dualising \eqref{NS5DBIsplit} on two directions \cite{Chatzistavrakidis:2013jqa}, so $i=1,2$. Following our (slightly different) conventions with the T-duality rules of \eqref{2Ts}, the action takes the form:
\be
\begin{split} 
S_{DBI} & = T_{5} \int e^{-2 \phi} \det E 
\sqrt { 1 + \frac{e^{2\phi}}{\det E} (\cC_{12})^2 }
\\ & \qquad\qquad\times
\sqrt{ 
- \det \left(
g_{\mu \nu} \partial_\alpha X^\mu \partial_\beta X^\nu 
+ \frac{ \det g}{\det E} g^{ij}  D_\alpha \tilde Y_i D_\beta \tilde Y_j
-   \frac{ e^\phi  \mG_{\alpha \beta}  }{\sqrt {\det E + e^{2\phi} \cC_{12}^2 }}\right)
}  
\end{split}
\label{DBI522}
\ee
where $E_{ij} \equiv g_{ij} +B_{ij}$, $\cC_{12} = C_{12} + B_{12} C_0$, $D_\alpha \tilde Y_i \equiv \partial_\alpha \tilde Y_i + \partial_\alpha X^\mu A_{\mu i}$
and 
\be
\begin{split}
\mG_{\alpha \beta} & = 
2 \partial_{[\alpha} c_{\beta]} 
+ \left( \hat C_{\mu\nu 12} + B_{12} ( \hat C_{\mu\nu} - 2\hat C_{\mu i} A_\nu{}^i + \hat C_{ij} A_\mu{}^i A_\nu{}^j) \right) \partial_\alpha X^\mu \partial_\beta X^\nu 
\\ &  \qquad
- 2\epsilon^{ij} ( \hat C_{\mu i} - A_\mu{}^k \hat C_{ki} ) \partial_{[\alpha} X^\mu D_{\beta]} \tilde Y_j
- \left(
C_0 \epsilon^{ij} + \tilde B^{ij} ( C_{12} + B_{12} C_0 )
\right) D_{\alpha}\tilde Y_i D_{\beta} \tilde Y_j 
\end{split}
\ee
where $\epsilon^{12} = 1$ and
\be
\tilde B^{ij} = \frac{\det B}{\det E} (B_{ij})^{-1}\,.
\ee
This brane has two special isometry directions, and the corresponding coordinates $Y^i$ do not appear in the action -- instead, their duals $\tilde Y_i$ do.

A covariant form of this action, in terms of two Killing vectors, is provided in \cite{Kimura:2014upa}.

\subsection{Doubled formalisms} 
\label{DFTrev} 

We will now discuss the core ideas of the doubled formalisms that will allow us to rewrite the above duality orbit of brane actions as a single ${\rm O}(d,d)$ manifest action.
The first step, starting from a theory defined in terms of coordinates $(X^\mu,Y^i)$, is to introduce dual coordinates $\tilde Y_i$ such that the doubled coordinates $Y^M = (Y^i , \tilde Y_i)$ combine into an ${\rm O}(d,d)$ vector.
Treating coordinates and their duals on the same footing was used in \cite{Duff:1989tf,Tseytlin:1990nb,Tseytlin:1990va,Hull:2004in,Hull:2006va} to construct doubled worldsheet theories, in which the target space geometry is doubled. To avoid introducing extra degrees of freedom, the doubled coordinates must be chiral, obeying a constraint $d Y^M = S^M{}_N \star dY^N$, where $d$ and $\star$ are the worldsheet exterior derivative and Hodge star, and the matrix $S^M{}_N = ( \eta^{-1} \gM)^M{}_N$ squares to the identity. The corresponding spacetime theory is that of double field theory \cite{Siegel:1993th, Siegel:1993xq, Hull:2009mi, Hull:2009zb, Hohm:2010jy, Hohm:2010pp} (as in fact follows from the beta functional equations \cite{Berman:2007xn, Berman:2007yf, Copland:2011yh,Copland:2011wx}). 

The five-brane actions we will describe will make use of such doubled coordinates. Unlike the doubled worldsheet, we will not have an intrinsic worldvolume relationship between the coordinates and their duals. Rather we will posit a form of the doubled five-brane action which resembles the gauged sigma model action of the Kaluza-Klein monopole, in which half of the coordinates $Y^M$ will not appear, assuming the existence of $d$ (generalised) Killing vectors corresponding to (generalised) special isometry directions. 

\subsubsection*{${\rm O}(10,10)$ DFT}

The standard version of DFT involves doubling all coordinates in spacetime. 
The only bosonic fields are the generalised metric, $\gM_{MN}$, generalised dilaton, $d$, and RR spinor $\mathcal{C}$. 
The NSNS sector fields are contained in the the former two as in \eqref{genmet} and \eqref{gendil}. 
One convenient way to construct the RR spinor is as follows \cite{Hohm:2011dv}.
Introduce $d$ fermionic creation operators $\psi^i$ and $d$ fermionic annihilation operators $\psi_i$ obeying
\be
\{ \psi_i, \psi^j \} = \delta_i^j \quad,\quad
\{ \psi_i , \psi_j \} = \{ \psi^i , \psi^j \} = 0 \,.
\ee
The vacuum $\vac$ obeys $\psi_i \vac = 0$, and a general spinor has the form
\be
\lambda = \sum_p \frac{1}{p!} \lambda_{i_1 \dots i_p} \psi^{i_1} \dots \psi^{i_p} \vac \,,
\ee
where we work in Majorana representations ($\lambda_{i_1 \dots i_p}$ real). One can restrict to Majorana-Weyl spinors: spinors formed using only odd and even numbers of creation operators have opposite chirality. 
(The states of definite chirality are eigenspinors of $(-1)^{N_F}$ where the number operator is $N_F = \sum_i \psi^i \psi_i$.)

The gamma matrices obeying $ \{ \Gamma_M , \Gamma_N \} = \eta_{MN}$ can be defined by
\be
\Gamma_M = ( \sqrt{2} \psi_i , \sqrt{2} \psi^i ) \,.
\ee
We will raise the index on $\Gamma_M$ using $\eta^{MN}$. 
We define 
\be
\Gamma_{M_1 \dots M_n} = \Gamma_{[M_1} \dots \Gamma_{M_n]} \,.
\ee
The charge conjugation matrix is
\be
C = \begin{cases} 
C_+ \equiv \prod_i ( \psi^i + \psi_i ) & d \,\,\mathrm{odd} \\
C_- \equiv \prod_i ( \psi^i - \psi_i ) & d \,\,\mathrm{even}
\end{cases}
\ee
and we define the conjugate spinor by $\bar \lambda \equiv \lambda^\dagger C$ (where $(\psi^i)^\dagger = \psi_i$ and $(\psi_i)^\dagger = \psi^i$). 

The RR spinor of the $O(10,10)$ DFT is denoted by $\mathcal{C}$ and has components 
\be
\mathcal{C}_{i_1 \dots i_p} = [ e^{B_2} C]_{i_1 \dots i_p}
\label{RRdict}
\ee
where $C$ is a polyform of RR potentials, described in Appendix \ref{RRconv}. This will be a chiral spinor, with chirality depending on whether we are in a IIA or IIB frame. 
A Buscher duality in the $i$ direction follows from acting with $\psi^i + \psi_i$, and changes chirality.

Let us note also that the gauge symmetries are (using $\eta_{MN}$ to raise and lower indices) 
\be
\begin{split} 
\delta_\Lambda \gM_{MN} &  = \Lambda^P \partial_P \gM_{MN} + 2 \partial_{(M} \Lambda^P \gM_{N)P} - 2 \partial^P \Lambda_{(M} \gM_{N) P}\,, \\
\delta_\Lambda e^{-2d} & = \partial_P ( \Lambda^P e^{-2d}) \,,\\
\delta_{\Lambda,\lambda} \cC & = \Lambda^N \partial_N \cC + \frac{1}{2} \partial_M \Lambda_N\Gamma^M \Gamma^N \cC + \slashed{\partial} \lambda \,,
\end{split} 
\label{DFTgauge}
\ee
where $\Lambda^M = ( \Lambda^i, \tilde \Lambda_i)$ encodes diffeomorphisms and gauge transformations of the B-field, while $\lambda$ is another spinor and gives to gauge transformations of the RR fields. The slashed partial derivative is
\be
\slashed{\partial} \equiv \frac{1}{\sqrt{2}} \Gamma^M \partial_M = \psi^M \partial_M \,,
\ee
where $\psi^M = \frac{1}{\sqrt{2}} \Gamma^M = (\psi^i ,\psi_i)$. 

\subsubsection*{${\rm O}(d,d)$ DFT}

Alternatively, we may choose to only double a subset $d < 10$ of the coordinates, along the lines of \cite{Hohm:2013nja}. 
This produces a slightly more intricate structure. 
The DFT coordinates are now $(X^\mu , Y^M)$.
The bosonic fields are the external metric, $g_{\mu\nu}$, the generalised metric, $\gM_{MN}$, and generalised dilaton, $d$, as well as a tensor hierarchy consisting of a one-form, $A_\mu{}^M$ and two-form, $B_{\mu\nu}$. The dictionary relating these to the 10-dimensional supergravity fields is the same as that presented in equations \eqref{SUGRAdecomp} to \eqref{gbdsplit} (so it is the same as one would use in Kaluza-Klein reduction, except one does not assume any coordinate independence). In addition, one can include RR potentials, $\mathcal{C}, \mathcal{C}_{\mu}$, $\mathcal{C}_{\mu\nu},\dots$ which are ${\rm O}(d,d)$ spinors of opposite chirality for fields with even or odd numbers of external indices. 
These RR potentials, by decomposing \eqref{RRdict}, correspond to
\be
\mathcal{C}_{\mu_1 \dots \mu_n i_1 \dots i_p } = [ e^{\hat B_2} \wedge \hat C ]_{\mu_1 \dots \mu_n i_1 \dots i_p} \,.
\ee
The gauge symmetries can be obtained by decomposing \eqref{DFTgauge}, first letting $\hat M$ be the ${\rm O}(10,10)$ index and then taking $\Lambda^{\hat M} = ( \xi^\mu , \tilde \Lambda_\mu , \Lambda^M )$
(in \cite{Hohm:2013nja} the component $\tilde \Lambda_\mu$ is taken to have the opposite sign).
In particular, under the gauge transformations $\tilde \Lambda_\mu$ one has
\be
\begin{split}
\delta A_\mu{}^M & = - \partial^M \tilde \Lambda_\mu \,, \\
\delta B_{\mu\nu} &= 2 \partial_{[\mu} \tilde \Lambda_{\nu]} - A_{[\mu}{}^M \partial_M \tilde \Lambda_{\nu]} \,,\\
\delta \cC_{\mu_1 \dots \mu_n} & = n (n-1) \partial_{[\mu_1} \tilde \Lambda_{\mu_2} \cC_{\mu_3 \dots \mu_n ]} 
 + (-1)^n n \frac{1}{\sqrt{2}} \Gamma^M \partial_M \tilde \Lambda_{[\mu_1} \cC_{\mu_2 \dots \mu_n ]}\,.
\end{split} 
\label{splitgauge}
\ee
One can make these transformations look nicer by writing them in a ``covariant'' form as in \cite{Hohm:2013nja}, however we will not do this here. 
For the RR spinors, one has also have transformations under external diffeomorphisms
\be
\delta_\xi \cC_{\mu_1 \dots \mu_n } = \xi^\nu \partial_\nu \cC_{\mu_1 \dots \mu_n} + n \partial_{[\mu_1} \xi^\nu \cC_{|\nu| \mu_2 \dots \mu_n]}
+ (-1)^{n-1} \frac{1}{\sqrt{2}} \Gamma^M \partial_M \xi^\nu \cC_{\nu \mu_1 \dots \mu_n }
\ee
and under RR gauge transformations:
\be
\delta_\lambda \cC_{\mu_1 \dots \mu_n} =  n \partial_{[\mu_1} \lambda_{\mu_2 \dots \mu_n]} + (-1)^n \frac{1}{\sqrt{2}} \Gamma^M \partial_M \lambda_{\mu_1 \dots \mu_n} \,,
\ee
while their transformation under generalised diffeomorphisms $\Lambda^M$ has the same form as before.

\section{O$(d,d)$ covariant DBI action}
\label{ODDDBI}

\subsection{The action}

\subsubsection*{Building blocks} 

We can write an ${\rm O}(d,d)$ covariant form of five-brane actions. 
The coordinates that appear in the action as worldvolume scalars are $(X^\mu, Y^M)$, where we have $n$ undoubled coordinates $X^\mu$ and the $2d$ doubled coordinates $Y^M$.
We introduce the ${\rm O}(d,d)$ generalised metric, $\gM_{MN}$, the ${\rm O}(d,d)$ one-form $A_\mu{}^M$, the external metric $g_{\mu\nu}$ and the generalised dilaton $e^{-2d}$.
We write a covariant differential for the doubled coordinates:
\be
D_\aw Y^M = \partial_\aw Y^M + \partial_\aw X^\mu A_\mu{}^M \,,
\ee
where $\aw,\bw$ are worldvolume indices. 

To describe the action for DFT monopoles, we adopt the techniques of \cite{Bergshoeff:1997gy, Eyras:1998hn} where the Kaluza-Klein monopole is viewed as a ``gauged sigma model.''
We need to introduce $d$ generalised Killing vectors, $k_a{}^M$ where $a=1,\dots,d$. 
(A generalised Killing vector is simply defined to annihilate the fields under the transformations $\delta_{k_a}$ corresponding to generalised diffeomorphisms. In adapted coordinates, we have as usual $k_a^M \partial_M = 0$ acting on all fields.)
These correspond to some special isometry directions, in a sense. 
Next define the matrix
\be
\hh_{ab} = \gM_{MN} k_a{}^M k_b{}^N \,,
\ee
with which one write projected (or ``gauged'') differentials
\be
\hat D_\alpha Y^M = D_\alpha Y^M - ( \hh^{-1} )^{ab}  k_a{}^M k_b{}^N \gM_{NP} D_\alpha Y^P \,,
\label{projdiff}
\ee
which will have the effect of removing half the doubled coordinates from the action. 
In order that the matrix $\hh_{ab}$ be invertible, we need
\be
T^{M_1 \dots M_d} \equiv \epsilon^{a_1 \dots a_d } k_{a_1}^{M_1} \dots k_{a_d}^{M_d} \,,
\label{T}
\ee
to be non-zero, as 
\be
\det \hh = \frac{1}{d!} \gM_{M_1 N_1} \dots \gM_{M_d N_d} T^{M_1 \dots M_d} T^{N_1 \dots N_d}
\ee
Later on, we will discuss how one can view this $T^{M_1 \dots M_d}$ as the T-duality covariant charge of the 5-brane, in line with the classification of \cite{Bergshoeff:2011zk}. 
In addition, we take 
\be
\eta_{MN} k_a^M k_b^N = 0 \,,
\label{knull}
\ee
which will in effect act as a sort of algebraic section condition on the worldvolume action. Different solutions of this constraint impose the existence of different special isometry directions in spacetime, and allow us to remove the corresponding scalar fields $Y^M$ from the brane worldvolume action that we will consider. 
Effectively, the condition \eqref{knull} implies that the $k_a^M$ live in an at most $d$-dimensional subspace, while requiring the object \eqref{T} be non-zero then implies that in fact they are a set of $d$ independent vectors. 

We also include the RR sector, introducing a set of forms which are ${\rm O}(d,d)$ spinors: $\mathcal{C}, \mathcal{C}_\mu, \mathcal{C}_{\mu\nu} , \dots$. 
Alongside the generalised Killing vectors, we have to introduce an auxiliary ${\rm O}(d,d)$ spinor $\lambdab$.
We require that it satisfy the following constraint:
\be
\Gamma_M \lambdab k^M_a = 0.
\ee
As there are $d$ independent $k_a^M$, this implies that $\lambdab$ is annihilated by half the ${\rm O}(d,d)$ gamma matrices $\Gamma^M$, and therefore it is a pure spinor. (Note that this then implies \eqref{knull}.)
 
Finally, we set the scale of $\lambdab$ by requiring that
\be
\frac{1}{( \sqrt{2} )^d }\blambdab \Gamma^{M_1 \dots M_d } \lambdab = T^{M_1 \dots M_d } \,.
\ee
This is the only non-zero spinor bilinear involving $\lambdab$ and its conjugate. 

\subsubsection*{The action} 

We may now write down the full DBI part of the action we consider:
\be
\begin{split}
S_{DBI} & = 
\int d^6 \sigma e^{-2d} \sqrt{ \det\hh } \sqrt{ 1 + e^{2d} ( \det \hh)^{-1/2} ( \blambdab \mathcal{C} )^2 }\times
\\ & 
\times \sqrt{ 
- \det \left(
g_{\mu\nu} \partial_{\aw} X^\mu \partial_{\bw} X^\nu 	+ \gM_{MN} \hat D_{\aw} Y^M \hat D_{\bw} Y^N 
-
\frac{ e^d ( \det \hh)^{-1/4} \blambdab \mG_{\aw \bw}}{ \sqrt{ 1 + e^{2d} ( \det \hh)^{-1/2} ( \blambdab \mathcal{C} )^2 }}
\right)
}\,,
\end{split}
\label{fullaction}
\ee
where 
\be
\mG_{\alpha \beta} = 2 \partial_{[\alpha} \cc_{\beta]} + \widetilde{\mathcal{C}}_{\alpha \beta}
\ee
is a worldvolume field strength with the following pullback of RR fields:
\be
\begin{split} 
\widetilde{\mathcal{C}}_{\aw \bw}
& = 
\left( 
\mathcal{C}_{\mu\nu} - ( B_{\mu\nu} + \frac{1}{2} A_\mu{}^M A_\nu{}^N \Gamma_{MN} ) \mathcal{C} 
+ \sqrt{2} A_\mu{}^M \Gamma_M \mathcal{C}_\nu  
\right) \partial_{[\aw} X^\mu \partial_{\bw]} X^\nu
\\ & 
\quad + 
\sqrt{2} \Gamma_M \left( \mathcal{C}_\mu - \frac{1}{\sqrt{2}} A_\mu{}^N \Gamma_N \mathcal{C} \right) \partial_{[\aw} X^\mu \hat D_{\bw]} Y^M 
\\ & 
\quad 
- \frac{1}{2} \Gamma_{MN} \mathcal{C} \hat D_{[\aw} Y^M \hat D_{\bw]} Y^N \,.
\end{split} 
\label{RR2pullback}
\ee
The worldvolume one-form $\cc_\alpha$ is here taken to also be an ${\rm O}(d,d)$ spinor. 
It is easy to check that the expression \eqref{RR2pullback} is invariant under gauge transformations of the external B-field, using the formula \eqref{splitgauge}. We will discuss its transformation properties under RR gauge transformations in Section \ref{gaugeRR}. 

The term
\be
e^{-2d} \sqrt{ \det\hh } 
\ee
provides the string coupling dependence: the generalised dilaton factor $e^{-2d}$ tells us that this brane indeed will have tension scaling as $g_s^{-2}$. 

One way of looking at the action \eqref{fullaction} is to think of it as a function of $d$, the number of doubled directions.\footnote{Not to be confused here with $d$, the generalised dilaton.}
When $d=0$, the fields that appear can be trivially identified with the usual spacetime ones: thus $g_{\mu\nu}$ is the full metric, $e^{-2d} \equiv e^{-2\phi}$ is the usual exponential of the dilaton, and $\mathcal{C}_{\mu\nu} \equiv C_{\mu\nu}$ and $\mathcal{C} \equiv C_0$ are the usual RR 2-form and 0-form.
Then setting $\det \hh = 1$ and $\blambdab =1$ we immediately see that what we have is the usual DBI action for the IIB NS5 brane. 

At the other extreme, $d=10$, we only have the generalised metric, generalised dilaton and a single ${\rm O}(10,10)$ spinor $\mathcal{C}$. We can replace $D_\alpha Y^M$ with $\partial_\alpha Y^M$, and one has
\be
\widetilde{\mathcal{C}}_{\aw \bw} = - \frac{1}{2} \Gamma_{MN} \mathcal{C} \hat \partial_{[\aw} Y^M \hat \partial_{\bw]} Y^N \,,
\ee
where we still project, $\hat \partial_\alpha Y^M \equiv \partial_\alpha Y^M  - ( \hh^{-1} )^{ab}  k_a{}^M k_b{}^N \gM_{NP} \partial_\alpha Y^P$. 

For $0 < d < 10$, the action interpolates between these two cases. One can choose $d$ to correspond to the number of actual isometry directions of the backgrounds, in which case the ${\rm O}(d,d)$ covariance is unbroken by the section condition solution, and one can view it directly as the ${\rm O}(d,d)$ T-duality group. 

We stress that the action \eqref{fullaction} is \emph{covariant} under ${\rm O}(d,d)$. Acting with some Buscher type transformation will map us to a different duality frame. In that frame, we view \eqref{fullaction} as providing the DBI action for a brane that is dual to the IIB NS5 brane. 

\subsection{Analysis of the NSNS terms}
\label{NSNSDBI} 

Let us focus on the action with the RR and worldvolume fields set to zero. It is
\be
\begin{split}
S_{DBI}\big|_{RR=0}  = 
\int d^6 \sigma 
e^{-2d} \sqrt{ \det\hh } \sqrt{ 
- \det \left(
g_{\mu\nu} \partial_{\aw} X^\mu \partial_{\bw} X^\nu 	+ \gM_{MN} \hat D_{\aw} Y^M \hat D_{\bw} Y^N 
\right)
}
\,,
\end{split}
\label{DBINSNS} 
\ee
Note that one can also write
\be
\gM_{MN} \hat D_\alpha Y^M \hat D_\beta Y^N = \Pi_{MN} D_\alpha Y^M D_\beta Y^N 
\ee
with
\be
\Pi_{MN} = \gM_{MN} - ( \hh^{-1})^{ab}  k_a^P k_b^Q  \gM_{MP} \gM_{NQ} \,.
\ee

\subsubsection*{Fully doubled: $d=10$}

We consider first the situation in which we have doubled all directions in spacetime. The action is simply
\be
S_{DBI}\big|_{RR=0} = \int d^6 \sigma 
e^{-2d} \sqrt{ \det\hh } \sqrt{ 
- \det \left(
\gM_{MN} \hat \partial_{\aw} Y^M \hat \partial_{\bw} Y^N \right)}
\,.
\label{10dNSNS}
\ee
The section condition is supposed to be $\partial_i \neq 0$, $\tilde \partial^i = 0$, so that the background fields may depend on the coordinates $Y^i$ but not the $\tilde Y_i$. 
As it stands, any of these may in principle appear in the action as worldvolume scalars. 
We will show in this section how one may remove the $\tilde Y_i$, in which case this action describes the IIB NS5.

Suppose we take the generalised Killing vectors $k_a{}^M$ to lie only in dual directions, i.e. $k_a^i = 0$ and $\tilde k_{ai} \neq 0$. 
Then $\hh_{ab} = \tilde k_{ai} \tilde k_{bj} g^{ij}$ and $\det \hh = ( \det \tilde k )^2 \det g^{-1}$, where we view $\tilde k_{ai}$ as a $10 \times 10$ matrix and take its determinant. 
We find due to this that
\be
e^{-2d} \sqrt{ \det \hh} = e^{-2\phi} | \det \tilde k|\,.
\ee
Now, we have
\be
\hat \partial_\alpha Y^i = \partial_\alpha Y^i
\ee
and
\be
\hat \partial_\alpha Y_i = \partial_\alpha Y_i 
- (\hh^{-1})^{ab} \tilde k_{ai} \tilde k_{bj} g^{jk} \partial_\alpha Y_k
- (\hh^{-1})^{ab} \tilde k_{ai} \tilde k_{bj} B_{kl} g^{lj} \partial_\alpha Y^k
\ee
In adapted dual coordinates, where the components of the Killing vectors are given by $\tilde k_{ai} = \delta_{ai}$, one has $\tilde k_{ai} \tilde k_{bj} (\hh^{-1})^{ab} =  g_{ij}$, which is always true for $d$ independent vectors,  and so one finds
\be
\hat \partial_\alpha Y_i = B_{ij} \partial_\alpha Y^j \,.
\ee
As promised, half the coordinates -- in this case, those that are the duals in the section with physical coordinates $Y^i$ -- have been projected out. 
The action \eqref{10dNSNS} becomes
\be
S_{DBI}\big|_{RR=0} = \int d^6 \sigma \,|\!\det \tilde k|
e^{-2\phi} \sqrt{ 
- \det \left(
g_{ij} \partial_\alpha Y^i \partial_\beta Y^j 
\right)} 
\,.
\ee
(Note that we have not written a tension prefactor, say $S_{DBI} = \widetilde{T}_5 \int d^6 \sigma ( \dots )$, but ideally one should absorb the leftover factor of $\det \tilde k$ into $\widetilde{T}_5$ and identify this with the original $T_5$. This should be kept in mind below.)

If we did not choose the $k_a^M$ to lie only in dual directions, we would obtain alternative forms of this action. 
If there is a spacetime isometry in the direction $i=z$, then for instance picking $k_a{}^z \neq 0$ but $\tilde k_{az} = 0$ for one $a$ would give us the action for a KKM in type IIA instead. 
To explore these possibilities, we will restrict to $d <10$. 
In particular,
to make contact with the known forms of the NS5, KKM and $5_2^2$ actions, it is convenient to specify to the case $d=2$. 

\subsubsection*{Partially doubled: d=2}

We again write the NSNS part of the DBI action:
\begin{equation}
S_{DBI}\big|_{RR=0}=\int \d^6 \sigma e^{-2d}\sqrt{\det \hh}\sqrt{-\det\Big(g_{\mu\nu}\dt_\alpha X^\mu \dt_\beta X^\nu +\P_{MN} D_\alpha Y^M D_\beta Y^N \Big)},
\end{equation}
where
\begin{equation}
\begin{aligned}
\hh_{ab}&=\mH_{MN}k_a^Mk_b^M\,,\\
\P_{MN}&=\mH_{MN}-\hh^{ab}\mH_{MP}\mH_{NQ}k_a^P k_b^Q\,,\\
D_\alpha Y^M &= \dt_\alpha Y^M + \dt_\alpha X^\mu A_\mu{}^M\,.
\end{aligned}
\end{equation}
We have two generalised Killing vectors $k_a^M$ obeying $\eta_{MN} k_a{}^M k_b{}^N =0$.
There are three choices of solutions each leading to a different effective action
\begin{equation}
\begin{aligned}
& {\rm NS5}:&& k_a^M=(0, \tk_{am})\,,\\
& {\rm KKM}:&& k_a^M= \{(k_1^m,0), (0,\tk_{2m})\}\,,\\
& 5_2^2:&& k_a^M=(k_{am},0)\,.
\end{aligned}
\end{equation}
Since we are in the ${\rm O}(2,2)$ theory, these are the only choices (up to a diffeomorphism in the KK5 case) for the Killing vectors. For larger groups the orbit will become longer and include e.g. $5_2^3$ and $5_2^4$ branes.

We will consider the available possibilities case by case.
In doing so, we will make use of the dictionary in Section \ref{DFTrev} relating the components of the DFT fields $e^{-2d}$, $\gM_{MN}$, $A_\mu{}^M$ to the decomposition of the 10-dimensional fields $\hat g_{\hmu \hnu} = ( g_{\mu\nu}, A_\mu{}^i, g_{ij})$, $\hat B_{\hmu \hnu} = ( B_{\mu\nu} , A_{\mu i} , B_{ij})$. Here $\hmu$ is the original ten-dimensional index, and $i=1,\dots,d$ denotes the directions which are doubled.

\subsubsection*{NS5-brane}

Choosing $k_a^M=(0, \tk_{ai})$ we have $\hh^{ab}\tk_{ai}\tk_{ai}=g_{ij}$ 
so that the only non-vanishing component of the projected generalised metric is:
\begin{equation}
\begin{aligned}
\P_{ij}&=\mH_{ij}- \hh^{ab}\mH_{i}{}^k\mH_{j}{}^l\tk_{ak}\tk_{bl}\\
&=g_{ij}-B_{ik}g^{kl}B_{lj}-B_i{}^k B_j{}^l g_{kl}=g_{ij}
\end{aligned}
\end{equation}
Hence, for the worldvolume matrix whose determinant appears in the second square root in the action, we find 
\begin{equation}
g_{\mu\nu} \partial_\alpha X^\mu \partial_\beta X^\nu + g_{ij} 
( \partial_\alpha Y^i + \partial_\alpha X^\mu A_\mu{}^i ) 
( \partial_\beta Y^j + \partial_\alpha X^\mu A_\mu{}^j ) \,,
\end{equation}
which is just the usual Kaluza-Klein-esque decomposition of the full expression\\ $\hat g_{\hmu \hnu} \partial_\alpha X^{\hmu} \partial_\beta X^{\hnu}$.
Note that all the dual coordinates disappear because the corresponding components of the projected generalised metric $\P$ vanish. Alternatively, one could calculate:
\be
\hat D_\alpha Y^M = \begin{pmatrix} D_\alpha Y^i \\ B_{ij} D_\alpha Y^j \end{pmatrix} \,.
\ee
One also computes the determinant 
\begin{equation}
\det{h_{ab}}=\fr{1}{\det g_{ij}} (\det{\tk_{ai}})^2.
\end{equation}
so that for the NSNS only part of the DBI action one gets
\begin{equation}
S^{NS5}_{DBI}\Big|_{RR=0}=\int d^6 \sigma \,|\!\det \tilde k|
e^{-2\phi}\sqrt{-\det\Big(\hat g_{\hmu\hnu} \partial_\alpha X^{\hmu} \partial_\beta X^{\hnu}\Big) } \,,
\end{equation}
Since the Killing vectors can be chosen to be some constants, they can be moved out from the integration as an overall prefactor.

\subsubsection*{KK monopole}

We now turn to the Kaluza-Klein monopole.
We pick the $k_a{}^M$ so that we have one non-vanishing Killing vector in the geometric directions and one in the dual directions. The algebraic section condition enforces them to be $k_1^M = ( k_1^i,0)$ and $k_2^M = (0,\tilde k_{2i})$ with $k_1^i \tilde k_{2i} = 0$. We can take a representative solution to be:
\begin{equation}
\begin{aligned}
k_1^M&=(k_1^1,0,0) \equiv (k,0,0,0),\\
k_2^M&=(0,0,0, \tk_{22}) \equiv (0,0,0,\tilde k).
\end{aligned}
\end{equation}
It is important to realise that the section condition solution is still such that $(X^\mu, Y^i)$ define the physical spacetime and $\tilde Y_i$ are duals. 
However, we will see that this choice of the $k_a^M$ in fact removes the $Y^1$ coordinate from the brane action, and in its place its dual $\tilde Y_1$ appears.
With the $Y^1$ direction corresponding to an isometry, this can be used to see that the action obtain is as expected the T-dual of the NS5 brane action on the $i=1$ direction. 

The matrix $\hh_{ab}$ is found to be
\be
\hh_{ab} = \begin{pmatrix} 
k^2 ( g_{11} + g^{22} ( B_{12})^2) & k\tilde k B_{12} g^{22} \\
k\tilde k B_{12} g^{22} & \tilde k^2 g^{22} 
\end{pmatrix} 
\Rightarrow \det \hh = \frac{ ( k \tilde k g_{11} )^2}{\det g} \,.
\ee
Then one has 
\be
\hh^{ab} k_a{}^M k_b{}^N 
= \frac{1}{g_{11}}
\begin{pmatrix}
1 & 0 & 0 & - B_{12} \\
0 & 0 & 0 & 0 \\
0 & 0 & 0 & 0 \\
- B_{12} & 0 & 0 & \det g + (B_{12})^2 
\end{pmatrix} \,,
\ee
from which one gets 
\be
\Pi_{MN} = \frac{1}{g_{11}} \begin{pmatrix} 
0 & 0 & 0 & 0 \\
0 & \det g + (B_{12})^2 & - B_{12} & 0 \\
0 & - B_{12} & 1 & 0 \\
0 & 0 & 0 & 0
\end{pmatrix} 
\ee
or equivalently
\be
\hat D_\alpha Y^M = 
\begin{pmatrix} 
- \frac{g_{12}}{g_{11}} D_\alpha Y^2 \\ D_\alpha Y^2 \\ D_\alpha \tilde Y_1 \\ \frac{g_{12}}{g_{11}} D_\alpha \tilde Y_1 
\end{pmatrix} 
\ee
and hence the derivatives $D_\alpha \tilde Y_2$ and $D_\alpha Y^1$ do not appear in the action, leaving only $D_\alpha Y^2$ and $D_\alpha \tilde Y^1$.

Substituting all these expressions into the covariant effective action one obtains
\begin{equation}
\begin{aligned}
S_{DBI}\big|_{RR=0} 
  =& \
\int d^6 \sigma k \tilde k g_{11} e^{-2\phi} \times\\
&\times\sqrt{ -\det\left(
g_{\mu\nu} \partial_\alpha X^\mu \partial_\beta X^\nu + \frac{\det g}{g_{11}} D_\alpha Y^2 D_\beta Y^2 
+ \frac{1}{g_{11}} \tilde D_\alpha \tilde Y_1 \tilde D_\beta \tilde Y_1 
\right) 
} 
\end{aligned} 
\label{KKMaction1}
\end{equation}
where
\be
\tilde D_\alpha \tilde Y_1 \equiv \partial_\alpha \tilde Y_1 + \partial_\alpha X^\mu A_{\mu 1} + D_\alpha Y^2 B_{21} \,.
\ee
Now, let us first show this agrees with the known KKM action presented in Section \ref{5branerev}.
Using the usual Kaluza-Klein-esque decomposition of the 10-dimensional metric $\hat g_{\hmu \hnu}$ in Section \ref{DFTrev}, we find that the geometric piece can be written as
\be
\begin{split}
g_{\mu\nu} \partial_\alpha X^\mu \partial_\beta X^\nu 
+ \frac{\det g}{g_{11}} D_\alpha Y^2 D_\beta Y^2 
& = \hat g_{\hmu \hnu} \hat \partial_\alpha X^{\hmu} \hat \partial_\beta X^{\hnu} 
\end{split} 
\label{KKgeo}
\ee
where $\hmu= (\mu,i)$ and 
\be
\hat \partial_\alpha X^{\hmu} = \partial_\alpha X^{\hmu} - \frac{1}{|\hat k|^2} \hat k^{\hmu} \hat k_{\hnu} \partial_\alpha X^{\hnu} \,,
\ee
where we introduce a Killing vector $\hat k$ such that $\hat k^1 = k$ and $|\hat k|^2 = \hat g_{11} k^2$. We have $\hat \partial_\alpha X^\mu = \partial_\alpha X^\mu$, $\hat \partial_\alpha Y^2 = \partial_\alpha Y^2$ and $\hat \partial_\alpha Y^1 = - \frac{1}{\hat g_{11}} ( \hat g_{12} \partial_\alpha Y^2 + \hat g_{\mu 1} \partial_\alpha X^\mu)$. Identifying $\hat g_{ij} = g_{ij}$ and $\hat g_{\mu i} = g_{ij} A_\mu{}^j$ leads to \eqref{KKgeo}.

The piece that is non-geometric can be written as
\be
\tilde D_\alpha \tilde Y_1 \equiv \partial_\alpha \tilde Y_1 + \partial_\alpha X^{\hmu} \hat B_{\hmu 1} \,,
\ee
using the identification $A_{\mu i} = \hat B_{\mu i} + A_\mu{}^i \hat B_{ij}$, $B_{ij} = \hat B_{ij}$, where $\hat B_{\hmu \hnu}$ is the 10-dimensional B-field. 
The determinant in the action therefore contains the term
\be
\frac{1}{g_{11}} \tilde D_\alpha \tilde Y_1 \tilde D_\beta \tilde Y_1 
= \frac{1}{|\hat k|^2} ( \partial_\alpha \tilde Y + \hat k ^{\nu} \partial_\alpha X^{\hmu} \hat B_{\hmu \hnu} )
 ( \partial_\beta \tilde Y + \hat k ^{\nu} \partial_\beta X^{\hmu} \hat B_{\hmu \hnu} ) \,.
\ee
Here we renamed $\tilde Y \equiv k \tilde Y_1$. 

Finally, we consider the prefactor $e^{-2\phi} k \tilde k g_{11} = e^{-2\phi} |\hat k|^2 \frac{\tilde k}{k}$. 
Up to the constant term $\tilde k/k$, this is the correct prefactor with the norm of the Killing vector $\hat k$ corresponding to the special isometry direction appearing explicitly. 
Hence the action agrees with that in \cite{Eyras:1998hn} (up to sign conventions for the B-field). 

Observe that the generalised Killing vector $k_1^M$ becomes in this frame the special Killing vector $\hat k$ of the KKM background.
The other generalised Killing vector $k_2^M$, which still points in the dual directions, does not have a geometric interpretation, and instead continues to play its former role of removing the second dual coordinate, $\tilde Y_2$, from the action.

Now let us comment on the T-duality relating this action to that of the NS5. 
Note that T-dualising along an isometry given by a Killing vector $\hat k$, one has (see the appendix of \cite{Kimura:2014upa}, for instance)
\be
e^{2 \tilde \phi} = \frac{1}{|\hat k|^2} e^{2\phi} \,,
\ee
which accounts for how the dilaton in the NS5 frame transforms: $e^{-2\phi^{NS5}} = |\hat k|^2 e^{-2 \phi^{KKM}}$.
To analyse the rest of the action, let us simplify matters by assuming we are in adapted coordinates, where $k=1$. 
Then we can use the usual Buscher rules for a T-duality in the 1 direction:
\be
\begin{array}{ccl}
\tilde g_{11} & =&  \frac{1}{g_{11}} \\
\tilde g_{12} & =& - \frac{B_{12}}{g_{11}} \\
\tilde g_{22} & =& \frac{\det g + (B_{12})^2}{ g_{11}} \\
\tilde B_{12} & = & - \frac{g_{12}}{g_{11}} \\
\end{array}
\begin{array}{ccl}
\tilde A_{\mu 1} & = &A_\mu{}^1 \\
\tilde A_\mu{}^1 & = & A_{\mu 1} \\
 & & \\ & & \\ 
\end{array}
\ee
Written in terms of the dual quantities, the action in this frame has the form:
\be
\begin{split}
S_{DBI}\big|_{RR=0} 
  = 
\int d^6 \sigma  e^{-2\tilde \phi }
\sqrt{ -\det\left(
g_{\mu\nu} \partial_\alpha X^\mu \partial_\beta X^\nu 
+ \tilde g_{ij} D_\alpha  Y^i D_\beta Y^j 
\right) 
} 
\end{split} 
\label{KKMactiondual}
\ee
with $Y^i = ( \tilde Y^1 , Y^2)$ and $D_\alpha Y^i = \partial_\alpha Y^i + \partial_\alpha X^\mu \tilde A_\mu{}^i$.
This is nothing other than the NS5 brane action that we considered before. 

\subsubsection*{$5_2^2$-brane}

The final possibility is to take
\begin{equation}
\begin{aligned}
k_a^M&= ( k_a^i,0 ) \,.
\end{aligned}
\end{equation}
This will lead to the action of the $5_2^2$ brane. 
We proceed as before. We have the matrix
\be
\hh_{ab} = k_a^i k_b^j ( g_{ij} - B_{ik} g^{kl} B_{lj} )
= k_a^i k_b^j g_{ij}  \frac{ \det g + (B_{12})^2}{\det g} \,,
\ee
where the second equality is true because $d=2$.
Hence,
\be
\det \hh = ( \det k)^2 \frac{ ( \det g + (B_{12})^2 )^2}{ \det g}
\ee
where we take the determinant of $k_a^i$ viewed as a two-by-two matrix. From this it follows that 
\be
\hh^{ab} k_a^M k_b^N = \begin{pmatrix} \frac{\det g}{\det g + (B_{12})^2} g^{ij} & 0 \\ 0 & 0 \end{pmatrix} \,.
\ee
So one gets 
\be
\Pi_{MN} = \begin{pmatrix} 0 & 0 \\ 0 & \frac{\det g}{\det g + (B_{12})^2} g^{ij} \end{pmatrix} \,,
\ee
or 
\be
\hat D_\alpha Y^M = \begin{pmatrix} \frac{B_{12}}{ \det g + (B_{12})^2} \epsilon^{ij} D_\alpha \tilde Y_j \\ D_\alpha \tilde Y_i \end{pmatrix} \,.
\ee
As expected in this case, both physical coordinates $Y^i$ are projected out of the action, and their place is taken by two dual coordinates $\tilde Y_i$, which are viewed as extra worldvolume scalars in this frame. 
Letting $E_{ij} = g_{ij} +B_{ij}$, the action is written as
\be
S_{DBI}\big|_{RR=0}
= \int d^6\sigma e^{-2\phi} \,|\!\det k|\, |\! \det E| \sqrt{
-\det \left( g_{\mu\nu} \partial_\alpha X^\mu \partial_\beta X^\nu + \frac{\det g}{\det E} g^{ij} D_\alpha \tilde Y_i D_\beta \tilde Y_j \right)
}
\ee
which agrees with the corresponding part of the $5_2^2$ action derived in \cite{Chatzistavrakidis:2013jqa, Kimura:2014upa} by T-dualising the NS5 action on both directions $Y^i$, as presented in \eqref{DBI522} in Section \ref{5branerev}.

One could also work in an alternative parametrisation of the generalised metric, involving a bivector field $\beta^{ij}$ \footnote{For a comprehensive review of supergravity theory based on $\b$-formalism see \cite{Andriot:2013xca}.},
\be
\gM_{MN} = \begin{pmatrix} \tilde g_{ij} & \tilde g_{ik} \beta^{kj} \\ - \tilde \beta^{ik} g_{kj} & \tilde g^{ij} - \beta^{ik} \tilde g_{kl} \beta^{lj} \end{pmatrix}\,.
\ee
 Now all expressions become more compact and the calculations are identical to those for the NS5 brane -- which is of course because this choice of frame really expresses the generalised metric in terms of the dual variables.
We find for instance 
\begin{equation}
\begin{aligned}
\det h&=(\det k_a^i)^2 \det \tg,\\
\P^{ij}&=\tg^{ij}.
\end{aligned}
\end{equation}
Note that the dual dilaton is defined as $e^{-2d}=e^{-2\phi}/\sqrt{\tg}$. 
Hence, the effective action becomes
\begin{equation}
S_{DBI}^{5_2^2}=\int d^6 \sigma \,|\!\det k| e^{-2\phi} \sqrt{- \det \Big( g_{\mu\nu} \partial_\alpha X^\mu \partial_\beta X^\nu
+ \tilde g^{ij} D_\alpha \tilde Y_i  D_\beta \tilde Y_j \Big) }\,,	
\end{equation}
which can again be easily identified as the T-dual of the NS5 brane action.

Before moving on, let us comment on the T-duality monodromy that characterises this brane. 
This takes the form of a shift of the bivector, $\beta^{ij} \rightarrow \beta^{ij} + \Lambda^{ij}$. 
Acting on $k_a^M$, this is $k_a^i \rightarrow k_a^i + \Lambda^{ij} \tilde k_{aj}$, $\tilde k_{ai} \rightarrow \tilde k_{ai}$. Hence, as $\tilde k_{ai} = 0$ in this case, the $k_a^M$ are well-defined. (Actually, we could already have made this comment for the twice smeared NS5 brane, for which the monodromy appears as a shift of the B-field, $B_{ij} \rightarrow B_{ij} + \Lambda_{ij}$ for which $k_a^i \rightarrow k_a^i$, $\tilde k_{ai} \rightarrow \tilde k_{ai} + \Lambda_{ij} k_a^j$.)

\subsection{Analysis of the RR terms}

Recall from Section \ref{DFTrev} that the single ${\rm O}(d,d)$ spinor $\lambda$ is identified with a polyform $\sum_p \lambda_{(p)}$ in spacetime, and constructed using $d$ fermionic creation operators $\psi^i$, which together with the annihilation operators $\psi_i$ provide a representation of the ${\rm O}(d,d)$ Clifford algebra. 
Some useful results are that if $\lambda$ has components $\lambda_{i_1 \dots i_p}$, then
\be
\begin{split} 
(\psi^i \lambda)_{i_1 \dots i_p} & = p \delta^i_{[i_1} \lambda_{i_2 \dots i_p ]} \,,\\
(\psi_i \lambda)_{i_1 \dots i_p} & = \lambda_{i i_1 \dots i_p }\,,\\
\end{split} 
\ee 
while
\be
\begin{split} 
( \Gamma^{ij} \lambda )_{i_1 \dots i_p} &  = 2p (p-1) \delta^i_{[i_1} \delta^j_{i_2} \lambda_{i_3 \dots i_p]} \,,\\
( \Gamma_{ij} \lambda )_{i_1 \dots i_p} &  = - 2 \lambda_{ij i_1 \dots i_p} \,,\\
( \Gamma_{i}{}^j \lambda )_{i_1 \dots i_p} &  =  \delta_i^j \lambda_{i_1 \dots i_p} - 2 p \delta_{[i_1}^j \lambda_{|i| i_2 \dots i_p]} \,.\\
\end{split} 
\ee

\subsubsection*{Fully doubled: $d=10$}

In the ${\rm O}(10,10)$ frame which corresponds to the IIB NS5 brane, the pure spinor $\lambda$ is 
\be
\blambdab = ( \det \tilde k_{ai} )^{1/2} \psi^1 \dots \psi^{10} \vac 
\quad \Rightarrow
\blambdab = (\det \tilde k_{ai} )^{1/2} \langle 0 | \,.
\ee
The scale has been set after noting that $T^{M_1 \dots M_d}$ has non-zero component \\ $T_{i_1 \dots i_{10}} = \epsilon_{ i_1 \dots i_{10}} ( \det \tilde k_{ai} )$. 

There is a single ${\rm O}(10,10)$ spinor $\mathcal{C}$ whose components are (in the conventions of \cite{Hohm:2011dv} where the $B$-field is minus that of \cite{Chatzistavrakidis:2013jqa})
\be
\mathcal{C}_{i_1 \dots i_p} = [e^{B_2} \wedge C ]_{i_1 \dots i_p} 
\ee
where $C = C_0 + C_{2} + C_{4} + \dots$ is the sum of the RR forms in IIB. Thus $\mathcal{C}$ only contains even numbers of creation operators. 

We can easily compute the quantities that appear in the action \eqref{fullaction}. We have:
\be
\blambdab \mathcal{C} = ( \det \tilde k_{ai} )^{1/2} C_0 \,,
\ee
so that that $e^d ( \det h )^{1/4} (\blambdab \cC)^2 = e^{\phi} (C_0)^2$. Meanwhile, 
one can calculate 
\be
\begin{split} 
- \frac{1}{2} \frac{ \blambdab}{(\det \tilde k_{ai})^{1/2}} \Gamma_{MN} \mathcal{C} \hat \partial_\alpha Y^M \hat \partial_\beta Y^N 
& 
 = \mathcal{C}_{ij} \partial_\alpha Y^i \partial_\beta Y^j 
 - \mathcal{C}_{(0)} \partial_\alpha Y^i \hat \partial_\beta Y_i
 \\ & 
 = ( \mathcal{C}_{ij} - B_{ij} \mathcal{C}_{(0)}  ) \partial_\alpha Y^i \partial_\alpha Y^j \\
 & = C_{ij} \partial_\alpha Y^i \partial_\alpha Y^j \,,
\end{split} 
\ee
where we used adapted coordinates such that $\hat \partial_\alpha Y^i = \partial_\alpha Y^i$ and $\hat \partial_\alpha Y_i = B_{ij} \partial_\alpha Y^j$.
Thus the action reproduces the contributions of RR terms to the IIB NS5 DBI action.

\subsubsection*{Partially doubled: $0 < d < 10$}

In the ${\rm O}(d,d)$ frame which corresponds to the IIB NS5 brane, the pure spinor $\blambdab$ is
\be
\blambdab =  ( \det \tilde k_{ai} )^{1/2} \psi^1 \dots \psi^{D} \vac 
\quad \Rightarrow
 \blambdab = ( \det \tilde k_{ai} )^{1/2} \langle 0 | \,.
\ee
The DBI action now involves the three ${\rm O}(d,d)$ spinors $\mathcal{C}$, $\mathcal{C}_\mu, \mathcal{C}_{\mu\nu}$. 
We have that $\mathcal{C}$ and $\mathcal{C}_{\mu\nu}$ are formed from even numbers of creation operators, while $\mathcal{C}_\mu$ is formed from odd numbers.
The components of these spinors are just
\be
\begin{split} 
\mathcal{C}_{i_1 \dots i_p} & = [e^{\hat B_2} \wedge \hat C ]_{i_1 \dots i_p}\,, \\
\mathcal{C}_{\mu i_1 \dots i_p} & = [e^{\hat B_2} \wedge \hat C ]_{\mu i_1 \dots i_p} \,,\\
\mathcal{C}_{\mu\nu i_1 \dots i_p} & = [e^{\hat B_2} \wedge \hat C ]_{\mu\nu i_1\dots i_p} \,,
\end{split} 
\label{splitRRdict}
\ee
where we now denote the 10-d fields as $\hat B_{\hmu \hnu}$, $\hat C_0$, $\hat C_{\hmu\hnu}, \dots$ in order to make the connection with the DFT variables clearer after splitting $\hmu = (\mu,i)$.

Clearly, we still have
\be
\blambdab \mathcal{C} = ( \det \tilde k_{ai} )^{1/2} C_0 \,,
\ee
while we need to compute $\blambdab \widetilde{\mathcal{C}}_{\aw \bw}$ with $\widetilde{\mathcal{C}}_{\alpha \beta}$ as in  \eqref{RR2pullback}.
We find
\be
\begin{split} 
\frac{\blambdab \widetilde{\mathcal{C}}_{\aw \bw}}{(\det \tilde k_{ai})^{1/2}}
& = 
\left( 
\mathcal{C}_{\mu\nu {(0)}} -  B_{\mu\nu} \mathcal{C}_{(0)}
+ A_\mu{}^i A_\nu{}^j \mathcal{C}_{ij} - A_{[\mu}{}^i A_{\nu] i} \mathcal{C}_{(0)} 
+ 2 A_{[\mu}{}^i \mathcal{C}_{\nu] i} 
\right) \partial_{[\aw} X^\mu \partial_{\bw]} X^\nu
\\ & 
\quad + 
2\left( \mathcal{C}_{\mu i} -  A_{\mu i} \mathcal{C}_{(0)} - A_\mu{}^j \mathcal{C}_{ji} \right) \partial_{[\aw} X^\mu  D_{\bw]} Y^i
\\ & 
\quad 
+ ( \mathcal{C}_{ij} - B_{ij} \mathcal{C}_{(0)} ) D_{[\aw} Y^i \hat D_{\bw]} Y^j \,,
\end{split} 
\ee
after using $\hat D_\alpha Y_i = B_{ij} D_\alpha Y^j$.
Relating the components of the $B$-field as usual as
\be
\begin{split}
\hat B_{ij} & = B_{ij}\,, \\
\hat B_{\mu i} & = A_{\mu i} + A_\mu{}^j B_{ji}\,, \\
\hat B_{\mu\nu} &= B_{\mu\nu} - A_{[\mu}{}^i A_{\nu]i} + A_\mu{}^i A_\nu{}^j B_{ij} \,,
\end{split} 
\ee
we find that
\be
\begin{split} 
\frac{\blambdab \widetilde{\mathcal{C}}_{\aw \bw}}{(\det \tilde k_{ai})^{1/2}}
 & = 
\left( \hat C_{\mu\nu} - 2 \hat C_{\mu i} A_\nu{}^i + A_\mu{}^i A_\nu{}^j \hat C_{ij} \right) \partial_\alpha X^\mu \partial_\beta X^\nu 
\\ & \qquad
+ 2 ( \hat C_{\mu i} + C_{ij} A_\mu{}^j ) \partial_{[\alpha} X^\mu D_{\beta ] } Y^i 
+ \hat C_{ij} D_\alpha Y^i D_\beta Y^j
\\ & = \hat C_{\hmu \hnu} \partial_\alpha X^{\hmu} \partial_\beta X^{\hnu} 
\end{split} 
\label{RR2target}
\ee
so again the choice of $\lambda$ in this frame picks out the correct contribution of the RR fields to IIB NS5 DBI action, using $e^d ( \det h)^{-1/4} = e^\phi ( \det\tilde k_{ai} )^{-1/2}$.

As all quantities used here transform covariantly as ${\rm O}(d,d)$ spinors, we can transforming both $\lambdab$ and $\cC$ and obtain the correct expressions for the RR contributions to the DBI action in the KKM and $5_2^2$ cases.

\subsection{Charges}

The $d$ Killing vectors $k_a^M$ give rise to an antisymmetric charge
\be
T^{M_1 \dots M_d} = \epsilon^{a_1 \dots a_d} k_{a_1}^{M_1} \dots k_{a_d}^{M_d} \,.
\label{charge}
\ee
The determinant factor that appears in the action can be written as
\be
e^{-2d} \sqrt{ \det \hh}  =e^{-2d} \sqrt{ \frac{1}{D!} \gM_{M_1 N_1} \dots \gM_{M_d N_d} T^{M_1 \dots M_d} T^{N_1 \dots N_d}} \,.
\ee
The paper \cite{Bergshoeff:2011zk} analysed string solitons and their classification under T-duality, showing that they fall into totally antisymmetric representations of ${\rm O}(d,d)$.
In particular, the five-branes in $D=10-d$ dimensions appear in the antisymmetric representation with $d$ antisymmetric indices, and in fact this further splits into self-dual and anti-self-dual irreducible representations, corresponding to five-branes with vector multiplets (as for the IIB NS5 and its T-duals) and tensor multiplets (as for the IIA NS5 and its T-duals), respectively.

For instance, consider the case $d=2$. We require
\be
T^{MN} = \pm \frac{1}{2} \epsilon^{MN}{}_{PQ} T^{PQ} \,,
\ee
where $\epsilon^{MNPQ}$ is defined by $\epsilon^{12}{}_{12} = 1$, and indices are raised and lowered using $\eta_{MN}$. 
This leads to the following conditions:
\be
\begin{array}{cccc}
(+) & T^1{}_2 = T^2{}_1 = 0 & \,,& T^1{}_1 = T^2{}_2 \\
(-) & T^{12}  = T_{12} = 0 & \,, &T^1{}_1 = -T^2{}_2 \,.
\end{array}
\ee
We note that Buscher transformations have determinant $-1$ and so send $\epsilon_{MNPQ} \rightarrow - \epsilon_{MNPQ}$. This means that a charge that is self-dual in one frame will be anti-self-dual in another. 
However, in each case, the inequivalent representation will continue to describe the ``other'' five-brane duality chain. 
The duality orbits we are interested in therefore appear as in table \ref{Ttable}. 

\begin{table}[h]
\centering
\begin{tabular}{|c|c|c|c|c|c|c|}\hline
  & charge & & brane & charge & & brane \\\hline
 IIB & $T_{12}$ & $(+)$ & NS5 &  &  & \\
 IIA & $T^{1}{}_{2}$ & $(-)$ & KKM & $T_{12} $ & $(+)$ & NS5 \\
 IIB & $T^{12}$ & $(+)$ & $5_2^2$ & $T^1{}_2$ & $(-)$ & KKM \\
 IIA &  &  & & $T^{12}$ & $(+)$ & $5_2^2$\\\hline
\end{tabular}
\caption{Branes and their charges. Note that $T^2{}_1$ would also describe KKM.}
\label{Ttable}
\end{table} 

Note that T-duality in the direction $i$ acts by raising or lowering the index $i$. This suggests the charges $T^1{}_1$ and $T^2{}_2$ must lie in a different orbit entirely, as they are mapped back into themselves on Buscher transformations. However, the corresponding potentials do not correspond to BPS objects (they are in a non-supersymmetric conjugacy class according to the analysis of \cite{Bergshoeff:2011zk}, for instance). In fact, consistent with that these are not allowed by our condition $k_a^M k_b^N \eta_{MN} = 0$. Hence, one may understand this as the BPS condition.

\section{O$(d,d)$ covariant Wess-Zumino action}
\label{Section_WZ}

\subsection{Gauge transformations and worldvolume field strengths} 
\label{gaugeRR}

In the DBI part of the action, there were two types of contributions from the RR fields, reflecting the presence in the original NS5 DBI term of the RR 0-form, $C_0$, and the RR two-form, $ C_2$. 
Both of these fields appeared in a gauge invariant: the 0-form is trivially invariant, while the two-form appeared alongside a worldvolume one-form, $c_\alpha$, in the combination
\be
\mG_{\alpha \beta} = 2 \partial_{[\alpha} c_{\beta]} + \hat C_{\hmu \hnu} \partial_\alpha X^{\hmu} \partial_\beta X^{\hnu} \,,
\ee
where under gauge transformations $\delta C_{\hmu\hnu} = 2 \partial_{[\hmu} \hat \lambda_{\hnu]}$ we have $\delta c_\alpha = - \hat \lambda_{\hmu} \partial_\alpha X^{\hmu}$.

In the DFT reformulation, the situation is more complicated. 
Consider the case $d=10$. There is a single RR spinor $\cC$, carrying no (doubled) spacetime indices, and transforming under RR gauge transformations as $\delta \cC = \psi^M \partial_M \lambda$, where $\lambda$ is a spinor of opposite chirality to $\cC$ (and where again $\psi_M \equiv \frac{1}{\sqrt{2}} \Gamma_M = ( \psi_i , \psi^i)$).

We obtained the term involving $C_0^2$ which appears in the NS5 action by making use of the pure spinor $\lambdab$, which was defined up to scale by $k_a^M \Gamma_M \lambdab = 0$.  
But consider the gauge transformation
\be
\delta ( \blambdab \cC ) = \blambdab \psi^M \partial_M \lambda \,,
\ee
which is apparently non-zero, unless 
\be
\psi^M \lambdab \partial_M = 0 \,.
\ee
One can argue that this must in fact be true: in coordinates adapted to the isometry, we have $k_a^M = \delta_a^M$ and $\partial_a = 0$. 
Let $\tilde a$ denote the remaining $d$ directions which do not correspond to the isometries associated to $k_a^M$ (some of these directions may be dual directions, depending on the choice of section, in which case also nothing will depend on them, but we can ignore this possibility). The directions $a$ must be dual to the $\tilde a$. The definition of $\lambdab$ is that $\psi_{\tilde a} \lambdab = 0$. Then one sees that $\psi^a \lambdab \partial_a + \psi^{\tilde a} \lambdab \partial_{\tilde a} = 0$, where the first term is zero because $\partial_a = 0$ and the second term is zero by the definition of $\lambdab$.

Hence although $\cC$ itself is not gauge invariant, the pullback $\blambdab \cC$ appearing in the worldvolume action will be. 

Now we can move on to discuss the more complicated pullback \eqref{RR2pullback} which appeared in the determinant part of the DBI term. 
This involves the RR fields $\cC, \cC_\mu, \cC_{\mu\nu}$. These have the following RR gauge transformations: 
\be
\begin{split}
\delta \cC & = \psi^M \partial_M \lambda \,, \\ 
\delta \cC_\mu & = \partial_\mu \lambda - \psi^M \partial_M \lambda_\mu \,,\\
\delta \cC_{\mu\nu} & = 2 \partial_{[\mu} \lambda_{\nu]} + \psi^M \partial_M \lambda_{\mu\nu} \,,
\end{split}
\ee
where $\lambda_{\mu\nu}$ and $\lambda$ have the opposite chirality to $\cC_{\mu\nu}$ and $\cC$, while $\lambda_\mu$ has the same chirality (but opposite to $\cC_{\mu}$). 
The variation involving $\lambda_{\mu\nu}$ is simply: 
\be
\delta \widetilde \cC_{\alpha \beta} 
= 
\psi^M \partial_M \lambda_{\mu\nu} \partial_\alpha X^\mu \partial_\beta X^\nu \,,
\ee
while that involving $\lambda_\mu$ is:
\be
\delta \widetilde \cC_{\alpha \beta} 
 = 2 \partial_{[\alpha} ( \lambda_\mu \partial_{\beta]} X^\mu ) 
+ 2 \psi^N\psi_M \partial_N \lambda_\mu  \left( A_\mu{}^M  \partial_{[\alpha} X^\mu \partial_{\beta]} X^\nu
+  \partial_{[\alpha} X^\mu \hat D_{\beta]} Y^M \right) \,,
\ee
and that involving $\lambda$ is:
\be
\begin{split}
\delta \widetilde \cC_{\alpha \beta} 
& =
2 \psi_M \partial_{[\alpha} \lambda \hat \partial_{\beta]} Y^M 
 + 2 \partial_{[\alpha} \lambda \psi_M ( \hat A_{\beta]}{}^M - A_{\beta]}{}^M)
- 2 \hat D_{[\alpha} Y^N \psi_M \partial_N \lambda  ( \hat A_{\beta]}{}^M - A_{\beta]}{}^M )
\\ & \quad - \psi^P \left(  B_{\alpha \beta}
+ \psi_M \psi_N \left[
\hat D_{[\alpha} Y^M \hat D_{\beta]} Y^N + 2 A_{[\alpha}{}^M  \hat D_{\beta]} Y^N
+ A_{[\alpha}{}^M A_{\beta]}{}^N 
\right]
\right)\partial_P \lambda\,.
\end{split}
\ee
Here $\hat A_\mu{}^M  - A_\mu{}^M = - k_a^M k_b^N h^{ab} \gM_{MP} A_\mu{}^P$ and we have written worldvolume indices where we have contractions with $\partial_\alpha X^\mu$. 
These expressions seem quite strange. Notice though, that contracting with $\blambdab$ we find simply
\be
\delta ( \blambdab \widetilde \cC_{\alpha \beta} ) 
= 2 \blambdab \partial_{[\alpha} \left( \lambda_\mu{} \partial_{\beta]} X^\mu 
+ \psi_M \lambda 
\partial_{\beta]} Y^M
\right) 
\ee
using $\blambdab \psi_M k_a^M  = 0$, $\blambdab \psi^M \partial_M = 0$ and also $\blambdab \psi_M \partial_\alpha ( k^M_a \dots ) = 0$ (as the possibility of a derivative hitting $k_a^M$ should not affect the definition of $\lambdab$).
Thus the combination $\blambdab \widetilde \cC_{\alpha \beta}$ transforms into a total derivative.

We can then combine the pullback with a worldvolume one-form $\cc_\alpha$ to produce a gauge invariant field strength. This $\cc_\alpha$ is also an ${\rm O}(d,d)$ spinor, and we define now
\be
\mG_{\alpha \beta} = 2 \partial_{[\alpha} \cc_{\beta]} + \widetilde \cC_{\alpha \beta} 
\ee
with 
\be
\delta \cc_\alpha = - \lambda_\mu \partial_\alpha X^\mu - \psi_M \lambda \partial_\alpha Y^M + \dots 
\ee
where the dots indicate that in principle one may have additional terms which vanish in $2\blambdab \partial_{[\alpha} c_{\beta]}$ thanks to the $\blambdab$,
such that $\blambdab \mG_{\alpha \beta}$ is gauge invariant. 

Evidently, one wants to proceed to construct additional gauge invariant pullbacks using $\blambdab$ and the other RR spinors.
Let us outline how this would be done generally in the ${\rm O}(10,10)$ case, as here we only have $\cC$ transforming under $\lambda$. 
Clearly the way to pullback this spinor to get a $p$-form on the worldvolume is to contract $\cC$ with 
$\blambdab \Gamma_{M_1 \dots M_p} \hat dY^{M_1}\wedge \dots\wedge \hat d Y^{M_p}$.
In particular, for the case of the NS5 brane (in adapted coordinates), where $\hat d \tilde Y_i = B_{ij} d Y^j$, one finds explicitly that
\be
 \frac{1}{(\sqrt{2})^p } \blambdab \Gamma_{M_1 \dots M_p } \mathcal{C} \hat \partial_{[\alpha_1 } Y^{M_1}  \dots  \hat  \partial_{\alpha_n]} Y^{M_p} 
= (-1)^{p(p-1)/2} 
 C_{i_1 \dots i_p} \partial_{[\alpha_1} Y^{i_1} \dots \partial_{\alpha_p]} Y^{i_p}\,,
\ee
where note $(-1)^{p(p-1)/2}$ is $+1$ for $p=0,1,4,5$ and $-1$ for $p=2,3,6$.

Now, we can calculate that 
\be
\begin{split} 
\delta &\left(  \frac{1}{(\sqrt{2})^p } \Gamma_{M_1 \dots M_p } \cC 
\hat \partial_{[\alpha_1} Y^{M_1} \dots \hat \partial_{\alpha_p]} Y^{M_p} \right)
  \\
 & \qquad = 
 p \partial_{[\alpha_p} \left( \psi_{M_1} \dots \psi_{M_{p-1}} \lambda
\hat \partial_{\alpha_1} Y^{M_1} \dots \hat \partial_{\alpha_{p-1}]} Y^{M_{p-1}} 
\right)
\\ & \qquad\qquad\qquad
 + \frac{1}{2} p ( p -1) (p-2) \eta_{M_1 M_2} \psi_{M_3} \dots \psi_{M_{p-1}} \partial_{[\alpha_p} \hat{\partial}_{\alpha_1} Y^{M_1} \dots \hat \partial_{\alpha_{p-1}]} Y^{M_{p-1}} 
\\ & \qquad\qquad\qquad
- p (p-1) \psi_{M_1} \dots \psi_{M_{p-1}}  \lambda \partial_{[\alpha_p } \hat{\partial}_{\alpha_1} Y^{M_1} 
\hat \partial_{\alpha_2} Y^{M_2} \dots \hat \partial_{\alpha_{p-1}]} Y^{M_{p-1}} 
\\ & \qquad\qquad\qquad 
+ (-1)^p \psi^N \psi_{M_1} \dots \psi_{M_p} \partial_N \lambda
\hat \partial_{[\alpha_1} Y^{M_1} \dots \hat \partial_{\alpha_p]} Y^{M_p} \,.
\end{split} 
\ee
The last two lines always vanish on contraction with $\blambdab$. 

Let us define
\be
\widetilde C_{\alpha_1 \dots \alpha_p} =(-1)^{p(p-1)/2} \frac{1}{2^{p/2}} \Gamma_{M_1 \dots M_p} \cC 
\hat \partial_{[\alpha_1} Y^{M_1} \dots \hat \partial_{\alpha_p]} Y^{M_p} 
\ee
and introduce worldvolume form fields (which are ${\rm O}(d,d)$ spinors) $\tilde c_{\alpha_1 \dots \alpha_p}$ transforming as
\be
\begin{split} 
\delta \tilde c_{\alpha_1 \dots \alpha_p} & = - (-1)^{p(p-1)/2} \frac{1}{2^{p/2}} \Gamma_{M_1 \dots M_p} 
\hat \partial_{[\alpha_1} Y^{M_1} \dots \hat \partial_{\alpha_p]} Y^{M_p} \\
 & =
- (-1)^{p(p-1)/2} \psi_{M_1} \dots \psi_{M_p} 
\hat \partial_{[\alpha_1} Y^{M_1} \dots \hat \partial_{\alpha_p]} Y^{M_p} \,.
\end{split}
\ee
Then the following ``field strength''
\be
\begin{split} 
\mG_{\alpha_1 \dots \alpha_p} 
= p \partial_{[\alpha_1} \cc_{\alpha_2 \dots \alpha_p]} 
 - \frac{1}{2} p (p-1) (p-2) \eta_{MN}  \partial_{[\alpha_{1}} \hat\partial_{\alpha_{2}} Y^M \hat\partial_{\alpha_3} Y^Nc_{\alpha_4 \dots \alpha_p]}
+ \widetilde C_{\alpha_1 \dots \alpha_p} 
\,, 
\end{split} 
\ee
is such that $\blambdab \mG_{\alpha_1 \dots \alpha_p}$ is gauge invariant. 

We can rewrite these in form notation: first letting
\be
\widetilde \cC_p \equiv
 (-1)^{p(p-1)/2} 
\frac{1}{p!} \frac{1}{(\sqrt{2})^p} \Gamma_{M_1 \dots M_p} \cC \hat dY^{M_1} \wedge \hat d Y^{M_p}
\ee
then we have 
\be
\widetilde \mG_p \equiv  d \tilde c_{p-1} - \frac{1}{2} \eta_{MN} d\hat dY^M \wedge \hat dY^N \wedge \tilde c_{p-3} 
+ \widetilde \cC_p\,,
\ee
and 
\be
\delta \cc_p = - (-1)^{p(p-1)/2} \frac{1}{p!} \frac{1}{(\sqrt{2})^p} \Gamma_{M_1 \dots M_p} \lambda \hat dY^{M_1}\wedge\dots\wedge \hat dY^{M_p} \,.
\ee
In the NS5 frame, we find that 
\be
\blambdab \widetilde \mG_p =\mG_p 
\ee
where
\be
\mG_p = dc_{p-1} + H_3 \wedge c_{p-3} + C_p
\label{wvolG}
\ee 
with $\delta c_p = - [ e^{-B_2} \lambda ]_p$, after identifying $\blambdab \cc_p = c_p$. The key point here is that the use of the projected coordinates allows us to obtain the $H_3$ factor in this frame from $d\hat dY^M$.

\subsection{Wess-Zumino term: RR contributions} 

As explained in Appendix \ref{RRconv}, the Wess-Zumino term for the NS5 brane in 10-dimensions can be written using the field strengths \eqref{wvolG} as
\be
L_{WZ} = B_6 + \frac{1}{2} \left( \mG_6 C_0 - \mG_4 \wedge C_2 + \mG_2 \wedge C_4 - \mG_0 C_6 \right) 
\ee
We will discuss the $B_6$ in the following subsections. Here, using the results from the previous subsection, it is trivial to express the remaining terms as:
\be
\begin{split} 
L_{WZ}  & \supset \frac{1}{2}
\big( 
 \blambdab \widetilde\mG_6 \blambdab \cC 
- \blambdab \widetilde\mG_4 \wedge \blambdab \widetilde\cC_2 
\\ & \qquad\qquad
+  \blambdab \widetilde\mG_2 \wedge \blambdab \widetilde\cC_4
- \blambdab \widetilde\mG_0 \wedge \blambdab \widetilde\cC_6 
\big)
\end{split} 
\ee
which is entirely ${\rm O}(10,10)$ covariant. This can be Fierzed into (now suppressing the wedge symbols for clarity)
\be
\begin{split}
L_{WZ}  & \supset  \frac{1}{2} \frac{1}{2^{10}}
\blambdab \Gamma^{M_1 \dots M_{10}} \lambdab
\big(
\widetilde\mG_6 \Gamma_{M_1 \dots M_{10}} \cC
-
\widetilde\mG_4 \Gamma_{M_1 \dots M_{10}} \widetilde\cC_2 
\\ & \qquad\qquad\qquad\qquad\qquad\qquad  +
\widetilde\mG_2 \Gamma_{M_1 \dots M_{10}} \widetilde\cC_4
-
\widetilde\mG_0 \Gamma_{M_1 \dots M_{10}} \widetilde\cC_6 
\big) \\
& =  \frac{1}{2} \frac{1}{(\sqrt{2})^{10}}
T^{M_1 \dots M_{10}}
\big(
\widetilde\mG_6 \Gamma_{M_1 \dots M_{10}} \cC
-
\widetilde\mG_4 \Gamma_{M_1 \dots M_{10}} \widetilde\cC_2 
\\ & \qquad\qquad\qquad\qquad\qquad\qquad  +
\widetilde\mG_2 \Gamma_{M_1 \dots M_{10}} \widetilde\cC_4
-
\widetilde\mG_0 \Gamma_{M_1 \dots M_{10}} \widetilde\cC_6 
\big) 
\end{split} 
\label{WZ10RR}
\ee
We claim that this represents part of an ${\rm O}(10,10)$ covariant Wess-Zumino term for the five-branes, and conjecture that reducing this to ${\rm O}(d,d)$ and imposing $\partial_M = 0$ should lead to the expressions in \cite{Bergshoeff:2011zk}.

\subsection{Wess-Zumino term: NSNS contributions}

We now turn to the leading term, which represents the magnetic potential to which the five-brane couples electrically. 
Matters are complicated here by the fact the duality orbit contains the Kaluza-Klein monopole, which couples to the magnetic dual of the Kaluza-Klein vector -- which is part of the metric. 
Let us now discuss some elements of how this is expected to appear in DFT. 
Recall that in ordinary supergravity, one can introduce $B_6$ as a Lagrange multiplier for the Bianchi identity for $B_2$. This 6-form field is sourced by the NS5-brane, whose T-dual KK monopole sources a field associated with the dual graviton. In the linear approximation this would be a vector-valued 7-form, i.e. 
\begin{equation}
\begin{aligned}
& B_{6}=B_{\hmu_1\dots \hmu_6} && \longleftrightarrow &&  A_{\hmu_1\dots \hmu_7,\hmu_8}=A_{7,1}.
\end{aligned}
\end{equation}
Further T-duality action generates fields $B_{8,2}$, $B_{8,3}$ and $B_{8,4}$ which interact with the $5_2^2,5_2^3$ and $5_2^4$ branes respectively. The latter is a co-dimension-0 object. Full classification of such objects in terms of irreps of ${\rm O}(d,d)$ can be found in \cite{Bergshoeff:2015cba,Bergshoeff:2011zk,Bergshoeff:2011ee}.

From the point of view of the full Double Field Theory, these potentials can be naturally associated with various Bianchi identities. Let us start with the split version of DFT which is formulated for a space of dimensions $D+2(10-D)$ parametrised by coordinates $(X^\mu,Y^M)$
and containing as before the fields $g_{\mu\nu}, A_{\mu}{}^M, B_{\mu\nu}, \gM_{MN}$ and $d$.
Consider first the form fields, $A_\mu{}^M$ and $B_{\mu\nu}$, which provide a ``tensor hierarchy'' similar to that found in gauged supergravities and in exceptional field theory.
The field strengths for the gauge potentials $A_\m{}^M$ and $B_{\m\n}$ satisfy  Bianchi identities of the following form
\begin{equation}
\label{BI_gauge}
\begin{aligned}
D_{[\m}\mF_{\n\r]}{}^M+\ldots + \partial^M \mH_{\mu\nu\rho} &=0,\\
D_{[\m}\mH_{\n\r\s]}+\ldots &=0,
\end{aligned}
\end{equation}
where the covariant derivative is defined in a Yang-Mills-like fashion $D_\m=\dt_\m-\mL_{A_\mu}$. 
In addition, the generalised metric contains components of the B-field and the metric for the internal space.
The corresponding field strength for these fields can be constructed by using the so-called flux formulation, defining generalised fluxes
$\mF_{MNK}$ and $\mF_M$ built using derivatives of the generalised vielbein, which themselves obey certain Bianchi identities (see Section \ref{Section_BI}). The corresponding Bianchi identities are simply
\be
\label{full_BI1}
 \dt_{[M}\mF_{NKL]}-\fr34\mF_{PMN}\mF_{QKL}\h^{PQ}=0\,.
\ee
Following the standard procedure as in \cite{Bergshoeff:2016ncb} one can introduce a dual potential $D_{MNKL}$ which acts as a Lagrange multiplier imposing the above identity when inserted into the DFT action. In principle one can perform this procedure for the identities \eqref{BI_gauge} obtaining dual potentials $D_{D-3,M}$ and $D_{D-4}$, where the number in the subscript  denotes the rank of the form in the external space. Note that at least for $D_{MNKL} $ this procedure only works at linear level due to the usual difficulties with the dual graviton.

Analysing tension for various objects of ${\rm O}(d,d)$-covariant theory living in $D=10-d$ dimensions in \cite{Bergshoeff:2011zk} it has been shown that magnetic gauge potentials  are the following
\begin{equation}
\begin{aligned}
&D_{D-4}, && D_{D-3,M},&& D_{D-2,MN}, && D_{D-1,MNK},   && D_{D,MNKL},  \\
& &&  && D_{D-2}, &&  D_{D-1,M},  && D_{D,MN},\\
& &&  &&  && && D_{D},
\end{aligned}
\end{equation}
where the $D-n$ subscript denotes rank of the form in the external $D$-dimensional space, and the ${\rm O}(d,d)$ indices are understand to be totally antisymmetric. As described above the top form potentials $D_{D,M_1\dots M_n}$ is related to the Bianchi identities for the fluxes $\mF_{MNK}$ and $\mF_{M}$ (at linear level), and the fields $D_{D-3,M}$ and $D_{D-4}$ could be related to the Bianchi identities for $\mF_2{}^M$ and $\mH_3$. The same procedure must naturally work for all other potentials, in particular the fields $D_{D-1,MNK}$ and $D_{D-1,M}$ generate ``mixed'' Bianchi identities via adding a Lagrange term of the following schematic form to the DFT action:
\begin{equation}
\e^{\m_1 \ldots \m_{D}}\Big((D_{[\m_1}\mF_{MNK}+\dots)D_{\m_2\ldots \m_D]}{}^{MNK}+(D_{[\m_1}\mF_{M}+\dots)D_{\m_2\dots \m_D]}{}^{M}\big).
\end{equation}
When considering a compactification ansatz these become the requirement that the embedding tensor is independent of the (external) coordinates. In principle one may start with the Bianchi identity \eqref{full_BI1} and understand that as the one formulated in the full O(10,10) theory. Then split of these identities upon $10=D+d$ will generate all known BI's from tensor hierarchy and many others, which correspond precisely to the potentials listed above.\footnote{Manifest demonstration of this procedure is work in progress and the results will be available soon.}

What we are interested in is the electric coupling of these dual potentials to the 5-branes.
Below we present explicit expressions first for $\rm{O}(10,10)$, and then ${\rm O}(2,2)$  and ${\rm O}(4,4)$ DFT which correspond to $D=8$ and $D=6$ respectively, essentially following the set-up of \cite{Bergshoeff:2011zk}, and indicating the choices of Killing vectors $k_a^M$ which pick out different branes in these cases.
Note that we are not precise about numerical factors, and will write $D_\alpha Y^M$ omitting the hat denoting the modification involving the generalised Killing vectors. 

In principle, one would want ultimately to fix the full Wess-Zumino term using gauge invariance. We can define the dual potential $D_{MNPQ}$ in linearised DFT \cite{Bergshoeff:2016ncb}, including its linearised NSNS gauge transformations, and this may be a good starting point. 
However, even if we do not know the full gauge transformations of the dual potentials (let alone how to define them non-linearly), we can and will proceed using the representation theoretic knowledge of what form they should take thanks to \cite{Bergshoeff:2011zk} and write down the only possible way they can couple to the 5-branes.

\subsubsection*{$D=0$ and O$(10,10)$} 

Firslty, let us consider the case where all direction are doubled. We know in this case that the charge has ten doubled indices, while the linearised dual of the generalised metric leads to a totally antisymmetric tensor with four indices, $D^{MNPQ}$ \cite{Bergshoeff:2016ncb}. We know that the worldvolume is six-dimensional, we also know that $6+4 = 10$, so it is natural to postulate in this case that  
\be
S_{WZ} \supset 
\int d\sigma^{\alpha^1} \wedge\dots\wedge d\sigma^{\alpha_6}  T_{ M_1 \dots M_{10}} D^{M_7 \dots M_{10} } \partial_{\alpha_1} Y^{M_1} \dots \partial_{\alpha_6} Y^{M_6} 
\label{WZ10NS} 
\ee
provides the coupling to the object $D^{MNPQ}$. 
In the NS5 frame, one has $T_{ i_1 \dots i_{10}} \sim \epsilon_{i_1 \dots i_{10}}$ and expect $D^{i_1 \dots i_4} \sim \epsilon^{i_1 \dots i_{10}} B_{i_5 \dots i_{10}}$, thereby automatically reproducing the expected $\int B_6$ term. 
The full WZ term, of course, should be given by combining \eqref{WZ10NS} with the RR contribution \eqref{WZ10RR}, which is justified here because we know that in the duality frame that corresponds to the NS5 brane, we obtain exactly the correct expected WZ term for the IIB NS5 brane (duality covariance effectively ensures that we should then obtain the right WZ terms for dual branes). 
It is tempting to wonder whether combining \eqref{WZ10NS} with the RR contribution \eqref{WZ10RR} sheds any light on the properties of $D^{MNPQ}$ -- for instance, partially fixing its RR gauge transformations.

\subsubsection*{$D=8$ and O$(2,2)$}

Consider next the case of ${\rm O}(2,2)$ DFT with 8 external coordinates $X^\m$, where one has the following potentials
\begin{equation}
\begin{aligned}
&D_{4}, && D_{5,M},&& D_{6,MN},  \\
& &&  && D_{6}.
\end{aligned}
\end{equation}
Note that one cannot have forms of rank larger than 6 as the worldvolume dimension of all the branes in question is 6. 

From the perspective of \emph{reduction} to $D=8$, we can think of each of these gauge potentials as coupling to differently embedded branes in the full ten-dimensional space split as $8+2$, which we can denote as follows:
\begin{equation}
\begin{aligned}
& 0 && 1 && 2 && 3 && 4 && 5 && 6 && 7 && | && 8 && 9 && &&\\
&\tm&&\tm&&\tm&&\tm&&   &&   &&   &&   && | &&\tm&&\tm&& D_{4}\\
&\tm&&\tm&&\tm&&\tm&&\tm&&   &&   &&   && | &&   &&\tm&& D_{5, M}\\
&\tm&&\tm&&\tm&&\tm&&\tm&&\tm&&   &&   && | &&   &&   && D_{6, MN}, && D_{6}\,.
\end{aligned}
\end{equation}
Hence, schematically one writes the following for the leading term in the Wess-Zumino action for the {\rm O}(2,2) theory on the DFT monopole  
\begin{equation}
\begin{aligned}
S_{WZ}&=\int T^{MN}d\s^{\a_1}\wedge \cdots \wedge d\s^{\a_6}\\
&\quad\times\Big( D_{\a_1\ldots \a_6 MN}+ D_{\a_1\ldots \a_5, M} D_{\a_6}Y_N+D_{\a_1\dots \a_4}D_{\a_5} Y_M D_{\a_6}Y_N\Big)\\
T^{MN}&=k_a^M k_b^N\e^{ab},
\end{aligned}
\end{equation}
where the integration is performed over the worldvolume of the brane and the three different terms involve the coordinates $Y^M$ which describe the fluctuations of the brane in the internal space - allowing for it to be wrapped in different orientations as in the scheme above.

To see explicitly that the above indeed reproduces the known Wess-Zumino terms for NS5, KK5 and $5_2^2$ branes, one first notes that the object $D_{\a_1\ldots \a_6 MN}$ combines the corresponding gauge potentials in an {\rm O}(2,2) covariant manner (here we write $V^M = (V^x,V^y, V_x, V_y)$)
\begin{equation}
\begin{aligned}
D_{\a_1\ldots \a_6}{}^{xy}&=B_{\a_1\ldots \a_6} && \rm NS5,\\
D_{\a_1\ldots \a_6 x}{}^y&=A_{\a_1\ldots \a_6 x,x} && \rm KK5,\\
D_{\a_1\ldots \a_6 y}{}^x&=A_{\a_1\ldots \a_6 y,y} && \rm KK5,\\
D_{\a_1\ldots \a_6 xy}&=D_{\a_1\ldots \a_6 xy,xy} && \rm 5_2^2.\\
\end{aligned}
\end{equation}
This implies, that these potentials are in the same representation as the charge and hence correspond to the same choice of the killing vectors $k_a^M$ solving the section constraint
\begin{equation}
\begin{aligned}
&\rm NS5:&& (0,\tk_{am})&& S_{WZ}=\tk_{1x}\tk_{2y}\e^{ab}\int B_{\a_1\ldots \a_6}d\s^{\a_1}\wedge \cdots \wedge d\s^{\a_6}+\dots,\\
&\rm KK5: &&  { \genfrac{}{}{0pt}{0}{(0,k_1^y,0,0)\hfill}{(0,0,\tk_{2x},0)} }
&& S_{WZ}=k_1^y\tk_{2x}\e^{12}\int A_{\a_1\ldots \a_6 y}{}^yd\s^{\a_1}\wedge \cdots \wedge d\s^{\a_6}+\dots,\\
&\rm KK5: &&  { \genfrac{}{}{0pt}{0}{(k_1^x,0,0,0)\hfill}{(0,0,0,\tk_{2y})} }
&& S_{WZ}=k_1^x\tk_{2y}\e^{12}\int A_{\a_1\ldots \a_6 x}{}^xd\s^{\a_1}\wedge \cdots \wedge d\s^{\a_6}+\dots,\\
&\rm 5_2^2: && (k_a^m,0) && S_{WZ}=k_a^x k_b^y\e^{ab}\int B_{\a_1\ldots \a_6 xy}{}^{xy}d\s^{\a_1}\wedge \cdots \wedge d\s^{\a_6}+\dots.
\end{aligned}
\end{equation}
We see, that the four classes of solutions of the constraint $k_a^Mk_b^N\h_{MN}=0$ correspond to the three branes with KK5 monopole combining two classes which differ only by $x\leftrightarrow y$.

\subsubsection*{$D=6$ and O$(4,4)$}

Consider now the case of ${\rm O}(4,4)$ DFT with 6 external coordinates $x^\m$, where one has the following potentials
\begin{equation}
\begin{aligned}
&D_{2}, && D_{3,M},&& D_{4,MN}, && D_{5,MNK},&& D_{6,MNKL},  \\
&       &&         && D_{4},    && D_{5,M},  && D_{6,MN},\\
&       &&         &&           &&           && D_{6},
\end{aligned}
\end{equation}
The corresponding possible embedding/wrapping table for the brane will look as follows
\begin{equation}
\begin{aligned}
& 0 && 1 && 2 && 3 && 4 && 5 && | && 6 && 7 && 8 && 9 && &&\\
&\tm&&\tm&&   &&   &&   &&   && | &&\tm&&\tm&&\tm&&\tm&& D_{2}, \\
&\tm&&\tm&&\tm&&   &&   &&   && | &&   &&\tm&&\tm&&\tm&& D_{3, M}\\
&\tm&&\tm&&\tm&&\tm&&   &&   && | &&   &&   &&\tm&&\tm&& D_{4, MN}, && D_{4}\\
&\tm&&\tm&&\tm&&\tm&&\tm&&   && | &&   &&   &&   &&\tm&& D_{5, MNK}, && D_{5,M}\\
&\tm&&\tm&&\tm&&\tm&&\tm&&\tm&& | &&   &&   &&   &&   && D_{6, MNKL}, && D_{6, MN}, D_{6}, D_{6}',
\end{aligned}
\end{equation}
note that we again do not list directions for the full doubled space leaving only half of them. As before the potentials in the last column do not corresponding to supersymmetric branes and hence cannot enter the Wess-Zumino term at the top level. 

Again schematic form of the Wess-Zumino term is simple and straightforward 
\begin{equation}
\label{WZ_s}
\begin{aligned}
S_{WZ}&= \int d\s^{\a_1}\wedge \cdots \wedge d\s^{\a_6}T^{MNKL}
\times\Big(D_{\a_1\ldots \a_6 MNKL} + D_{\a_1\ldots \a_5, MNK} D_{\a_6}Y_L \\
&+ D_{\a_1\ldots \a_4, MN} D_{\a_5}Y_K D_{\a_6}Y_L + D_{\a_1\a_2 \a_3, M}  D_{\a_4}Y_N D_{\a_5}Y_KD_{\a_6}Y_L \\
&+D_{\a_1\a_2} \; D_{\a_3}Y_M D_{\a_4}Y_N D_{\a_5}Y_KD_{\a_6}Y_L \Big),
\end{aligned}
\end{equation}
with the charge defined as
\begin{equation}
T^{MNKL}=k_a^Mk_b^Nk_c^Kk_d^L\e^{abcd}.
\end{equation}

Finding a general solution of the algebraic section constraint for {\rm O}(4,4) is a technically involved problem, however explicit computer check of various solutions shows that there are only 5 classes of solutions, which correspond precisely to $5_2^p$-branes with $p=0,1,2,3,4$, as they were found in \cite{Bakhmatov:2016kfn}. Let us list the representative solutions for these classes 
\begin{equation}
\label{preBI}
\begin{aligned}
&\rm 5_2^0:&& k_a^M=(\vec{0};\tk_{am})&& S_{WZ}=\tk_{1m}\tk_{2n}\tk_{3k}\tk_{4l}\e^{mnkl}\e_{1234}\int D_{6}{}^{1234}+\dots,\\ \\
&\rm 5_2^1: &&
\begin{aligned}
k_1^M&=(k_1^1,0,0,0;\vec{0})\\
k_{(2,3,4)}^M&=(\vec{0};0,\tk_{(2,3,4)\hat{m}})
\end{aligned} 
&& S_{WZ}=4k_1^m\tk_{2n}\tk_{3k}\tk_{4l}\e^{1nkl}\e_{m234}\int D_{6,1}{}^{234}+\dots,\\ \\
&\rm 5_2^2: && 
\begin{aligned}
k_{(1,2)}^M&=(k_{(1,2)}^{\tilde{m}},0,0;\vec{0})\\
k_{(3,4)}^M&=(\vec{0};0,0,\tk_{(3,4)\tilde{n}})
\end{aligned} 
&& S_{WZ}=4k_1^mk_2^n\tk_{3k}\tk_{4l}\e^{12kl}\e_{mn34}\int D_{6,12}{}^{34}+\dots.\\ \\
&\rm 5_2^3: && 
\begin{aligned}
k_{1}^M&=(\vec{0};\tk_{11},0,0,0)\\
k_{(2,3,4)}^M&=(0,k_{(2,3,4)}^{\hat{m}};\vec{0})
\end{aligned} 
&& S_{WZ}=4\tk_{1m}k_{2}^{n}k_{3}^{k}k_{4}^{l}\e^{m234}\e_{1nkl}\int D_{6,234}{}^{1}+\dots.\\ \\
&\rm 5_2^4: && k_a^M=(k_a^m,\vec{0}) && S_{WZ}=k_{1}^{m}k_{2}^{n}k_{3}^{k}k_{4}^{l}\e^{1234}\e_{mnkl}\int D_{6,1234}+\dots.
\end{aligned}
\end{equation}
where the normal and dual components of the generalised vectors $k_a^M$ are separated by semicolon for the sake of clarity. Just for this expression we define $\hat{m}=2,3,4$, $\tilde{m}=1,2$, $\tilde{n}=3,4$. Although the expression above might seem messy, the idea behind that is straightforward; one distinguishes five classes of solutions, depending on how many vectors have non-zero components in the geometric direction. 

One notes here, that the above expressions do not only reproduce the structure of the Bianchi identities for non-geometric backgrounds as obtained in \cite{Bakhmatov:2016kfn}, but also give the correct factor of 4 in the RHS. In addition, one observes the correct counting of indices of the gauge potential and the following relation between the potentials $D_{6,MNKL}$ and the gauge potentials $D_{6+p,p}$ which are associated to solutions with $p$ special directions
\begin{equation}
\begin{aligned}
D_{\m_1\dots\m_6}{}^{1234}&=B_{\m_1\dots\m_6},\\
D_{\m_1\dots\m_6,1}{}^{234}&=A_{\m_1\dots\m_6x,x},\\
D_{\m_1\dots\m_6,12}{}^{34}&=D_{\m_1\dots\m_6xy,xy},\\
D_{\m_1\dots\m_6,123}{}^{4}&=D_{\m_1\dots\m_6xyz,xyz},\\
D_{\m_1\dots\m_6,1234}&=D_{\m_1\dots\m_6xyzw,xyzw},
\end{aligned}
\end{equation}
with the obvious definition of the coordinates $x,y,z,w$.

Hence, one concludes, that the solutions listed in \cite{Bakhmatov:2016kfn} correspond to such embedding of the DFT monopole in the full {\rm O}(4,4) DFT when it interacts only with the 6-form potential $D_{6,MNKL}$. Such dependence of the solution on embedding is expected as the full 10D 6-form potential for the NS5-brane for example under the split $10=6+4$ gets decomposed to various forms. Covariantizing each of them under {\rm O}(4,4) one get the potentials $D_{p,M_1\dots M_m}$ listed above.

\subsection{Bianchi identities}
\label{Section_BI}

We now want to make some comments about the form of the Bianchi identities that are sourced by the five-brane action we have described. 
The Bianchi identities in the part of DFT described by the generalised metric and generalised dilaton can be formulated as follows.
First, we need to introduce a generalised vielbein, $E^A_M$, such that $\gM_{MN} = E^A_M E^B_N \gM_{AB}$, where the flat generalied metric $\gM_{AB}$ can be taken to be the identity if $d<10$, or by $\gM_{AB} = \mathrm{diag} (\bar\eta, \bar\eta^{-1} )$ where $\bar \eta$ is the flat Minkowski metric for $d=10$. The generalised vielbein is a group element and can be taken to obey $E^A_M E^B_N \eta^{MN} = \eta^{AB}$ where $\eta^{AB}$ will be chosen to be numerically equal to $\eta^{MN}$. 

From this generalised vielbein and its inverse, one can define the following generalised flux, given in flat indices as follows:
\begin{equation}
\begin{aligned}
\mF^{A}{}_{BC}&=2E^{A}_M E_{[B}^N\dt_N E_{C]}^M+E^{A}_M \h^{MN}\h_{KL}\dt_N E_{[B}^K E_{C]}^L,\\
\mF_A&=\dt_M E^M_A+2E^M_A\dt_Md.
\end{aligned}
\end{equation}
This flux identically satisfies the following Bianchi identities:
\begin{equation}
E^M_{[A}\dt_M\mF_{BCD]}-\fr34\mF^E{}_{[AB}\mF_{|E|CD]} \equiv \mZ_{ABCD}=0
\end{equation}
We can also define the flux in curved indices,
\begin{equation}
\mF_{MNK}=3E_{A[M}\dt_NE_{K]}{}^A\,,
\end{equation}
which now obeys the BI 
\be
\label{fullBI}
S_{MNPQ} \equiv \dt_{[M}\mF_{NKL]}-\fr34\mF_{PMN}\mF_{QKL}\h^{PQ}=0\,.
\ee
up to terms which vanish by the section condition.
The components of the generalised flux $\mF_{MNP}$ can be identified with a set of spacetime tensors: the three-form $H_{ijk}$, the ``geometric flux'' $\tau_{ij}{}^k$, the non-geometric flux, $Q_i{}^{jk}$ and the non-geometric R-flux, $R^{ijk}$.

In presence of sources the RHS of the Bianchi identities gets modified to include a delta function. Note that the harmonic function $H(y)$ which charactert an es a brane-like solution solves the Poincare equation with non-vanishing source term, which for the relevant discussion of \cite{Bakhmatov:2016kfn} reads
\begin{equation}
\sum_{i=1}^3\dt_i\dt_iH(y_1,y_2,y_3)=\a \d(y_1^2+y_2^2+y_3^2).
\end{equation}
Substituting the explicit expressions for the backgrounds from \cite{Bakhmatov:2016kfn} we obtain the following expression for the non-vanishing components of the Bianchi identities
\begin{equation}
\label{BI}
\begin{aligned}
5_2^0:&& S_{abcd}=\fr{\a}{4H^2}\e^{1234}\e_{abcd}\d\big((x^1)^2+(x^2)^2+(x^3)^2 \big),\\
5_2^1:&& S^a{}_{bcd}=\fr{\a}{H^2}\e^{a123}\e_{bcd4}\d\big((x^1)^2+(x^2)^2+(x^3)^2 \big),\\
5_2^2:&& S^{ab}{}_{cd}=\fr{\a}{H^2}\e^{ab12}\e_{cd34}\d\big((x^1)^2+(x^2)^2+(\tx^3)^2 \big),\\
5_2^3:&& S^{abc}{}_{d}=\fr{\a}{H^2}\e^{abc1}\e_{d234}\d\big((x^1)^2+(\tx^2)^2+(\tx^3)^2 \big),\\
5_2^4:&& S^{abcd}=\fr{\a}{4H^2}\e^{abcd}\e_{1234}\d\big((\tx^1)^2+(\tx^2)^2+(\tx^3)^2 \big),\\
\end{aligned}
\end{equation}
where we include the factors $\e^{1234}=1$ and $\e_{1234}=1$ for the sake of symmetry and $S_{MNKL}$ is just the full expression \eqref{fullBI}. Note that these precisely repeat the structure of \eqref{preBI}. Very similar if not the same Bianchi identities have been found in \cite{Andriot:2014uda}, where however appeared some inconsistency with the result of \cite{Chatzistavrakidis:2013jqa} (see Appendix D.2 of the former for more details). Hence, the result here and in \cite{Bakhmatov:2016kfn} are closer to the former, than the latter. 

 There is still the technical question of repeating of all the calculation of \cite{Bakhmatov:2016kfn} for the case of the localized DTF monopole, i.e. the one with $H=H(y_1,y_2,y_3,y_4)$. However, for sure this will not give new information and the whole discussion will just be repeated.

Taking into account these observation and the expression for the covariant WZ-action for the {\rm O}(4,4) theory we conjecture that the full covariant Bianchi identities should be of the following form
\begin{equation}
\dt_{[M}\mF_{NKL]}-\fr34 \mF_{P[MN}\mF^P{}_{KL]}\propto T_{MNKL}\delta\big(r^2(Y^M)\big).
\end{equation}
The function $r^2(Y)$ is always a sum of squares of 4 coordinates which are chosen by solving the differential section constraint.

\section{Discussion}
\label{discussion}

Let us recap the main accomplishments of this paper. 
We have introduced a formulation of the action for the IIB NS5 brane and its T-duals in the language of double field theory: this is an action for a brane in a doubled spacetime (with doubled coordinates $Y^M$ appearing as the worldvolume scalars), coupling to a background characterised by its generalised metric, ${\rm O}(d,d)$ NSNS tensor hierarchy form fields  and ${\rm O}(d,d)$ RR spinors. 
The full action for $d=10$ is given by the sum of the DBI action, \eqref{fullaction} (specialised to $d=10$), the WZ RR contributions \eqref{WZ10RR} and the WZ NS contribution \eqref{WZ10NS} in terms of the proposed dual field $D^{MNPQ}$.

The DFT solution describing these branes has been termed a ``DFT monopole''\cite{Berman:2014jsa}, while the exotic branes obtained by further T-dualities have been dubbed ``generalised monopoles'' in \cite{Bergshoeff:2011ee, Bergshoeff:2011mh}. Our doubled 5-brane action lives up to these expectations by mimicking the form of the action for the usual Kaluza-Klein monopole \cite{Bergshoeff:1997gy, Bergshoeff:1998ef, Eyras:1998hn}. We deal with having twice as many coordinates by treating half the doubled coordinates as corresponding to special isometry directions. Introducing (generalised) Killing vectors for these directions, we can construct a manifestly ${\rm O}(d,d)$ covariant action. These $d$ generalised Killing vectors $k_a^M$ allow us to construct a charge $T^{M_1 \dots M_d} = \epsilon^{a_1 \dots a_d} k_{a_1}^{M_1} \dots k_{a_d}^{M_d}$ characterising the 5-brane. They also allow us to define an auxiliary ${\rm O}(d,d)$ spinor $\lambdab$, which is needed to pullback the ${\rm O}(d,d)$ RR spinors to the brane worldvolume, by projecting out the components that should appear in the action different duality frames. These ingredients combine to produce the ${\rm O}(d,d)$ covariant DBI action, equation \eqref{fullaction}, while we provided the essential features of an ${\rm O}(d,d)$ covariant WZ term in Section \ref{Section_WZ}. 

This provides a unified formulation of a number of T-dual branes. One can obtain these dual formulations by making alternative choices of the $k_a^M$, while assuming in each case that the solution to the section condition $\eta^{MN} \partial_M \partial_N = 0$ is the same ($\tilde \partial^i = 0$) and that the coordinates $Y^i$ parametrise the physical spacetime while the $\tilde Y_i$ are dual. Then, as we saw, choosing a particular $k_a^M$ to correspond to an isometry in a dual direction ($k_a^M = ( 0 , \tilde k_{ai})$) or to one in a physical direction ($k_a^M = ( k_a^i,0)$) amounted to describing for instance the NS5 brane or the T-dual KKM. 
Note that to view this as an actual T-duality in the usual sense requires there to really be an isometry (corresponding to $k_a^i$). 
Otherwise, there is a somewhat subtle set of possibilities, which we will now discuss in detail. 

\subsection{Location, location, location}

Effectively, we have three choices: 
\begin{itemize}
\item the choice of solution of the usual ``differential'' section condition, $\eta^{MN} \partial_M \partial_N = 0$. The solution tells us which $d$ coordinates our fields and gauge parameters may depend on.
\item the choice of ``duality frame'', i.e. which $d$ coordinates are taken to be physical, i.e. to be those of the physical spacetime. In DFT, this does not have to be the same as the choice of solution to the section condition! Such a background would have physical coordinates $Y^i$ but may depend on the $\tilde Y_i$ as long as it did not also depend on the physical counterpart of any $\tilde Y_i$ that the fields depend on. Evidently, this is entirely non-geometric from the supergravity point of view and the general interpretation of such backgrounds in string theory is not clear. Clearly also this does not correspond to carrying out a ``duality'' in the usual sense. Let us refer to this instead as the choice of ``spacetime frame''.
\item the choice of the $d$ $k_a^M$, which correspond to the existence of $d$ isometries in the doubled spacetime. The differential section condition imposes that there are always $d$ isometries: these $k_a^M$ can correspond to isometries beyond the section condition. 
\end{itemize}

At the risk of overstating the point, let us consider a simple example in great detail.
Suppose we have a brane action for just two worldvolume scalars $(Y,\tilde Y)$, viewed as doubled coordinates, with some non-trivial background generalised metric $\gM(Y,\tilde Y)$ whose coordinate dependence is subject to the usual section condition. In principle, we have an action
\be
S = S[ Y , \tilde Y; \gM(Y, \tilde Y)]\,.
\ee
Then the possible choices we are faced with amount to the following:
\begin{itemize}
\item The section condition solution is $\partial_{\tilde Y} = 0$, the choice of ``spacetime frame'' is that $Y$ is the physical coordinate, and we have $k_a^M = (0,k)$. Then the action is $S = S[ Y ; \gM(Y) ]$, a wholly geometric action. This is analogous to the NS5 brane action. 
\item Alternatively, if we choose $k_a^M = (k,0)$ then the $Y$ direction is \emph{also} an isometry (as well as $\tilde Y$) but its derivatives are removed from the action using the gauged sigma model approach, and the action is $S = S[\tilde Y; \gM]$. This is mildly non-geometric in that the action is the action of a worldvolume scalar $\tilde Y$ (but the background is independent of both $Y$ and $\tilde Y$). This is analogous to the KKM and $5_2^2$ actions, and is T-dual to the subsequent case.
\item Now suppose we solve the section condition as $\partial_Y = 0$ and pick the ``spacetime frame'' such that $Y$ is the physical coordinate. The background may still depend on $\tilde Y$. This is what we would get by naively applying the Buscher rules along a direction which is not an isometry. 
If we take $k_a^M = (0,k)$ then we are forced into having $\tilde Y$ as an isometry direction also. The action is $S = S[ Y ; \gM ]$. This is analogous to the NS5 brane with an isometry, and is T-dual to the preceding case.
\item Instead take the previous case but with $k_a^M = (k,0)$. We still view $Y$ as a coordinate in spacetime, but the brane action is $S = S[ \tilde Y;\gM(\tilde Y)]$. This is as non-geometric as it gets for us. The background depends on dual coordinates: however we see that the fluctuations of the brane are in the $\tilde Y$ direction only and not in the physical spacetime $Y$. If this description can be trusted, this describes some an entirely locally non-geometric brane. The simplest example is maybe to think of this as describing the action for the KKM localised in winding space. 
The localisation of the KKM in this manner is expected from the worldsheet instanton calculation of \cite{Harvey:2005ab, Jensen:2011jna}.
\end{itemize} 
We see that operating within DFT provides us with the ability to choose the location of spacetime, the location of the coordinates that our background can depend on and, for the class of 5-branes considered in this paper, the location of further special isometry directions.

\subsection{Relation to other approaches and future work} 

Underlying our construction, was the technology of a ``gauged sigma model'', meaning that we required half the doubled directions be isometries, and we introduced generalised Killing vectors corresponding to these isometries which played a vital role in writing down the action. 
In practice, this involved modifying the derivatives of the worldvolume scalars corresponding to doubled coordinates, so that 
\be
\partial_\alpha Y^M \rightarrow \hat \partial_\alpha Y^M = \partial_\alpha Y^M - k_a^M (h^{-1})^{ab} k_b^N \gM_{NP} \partial_\alpha Y^P \,.
\ee
One could also view this as a gauging,
\be
\partial_\alpha Y^M \rightarrow \hat \partial_\alpha Y^M = \partial_\alpha Y^M + \gV_\alpha^M 
\ee
with a dependent gauge field $\gV_\alpha{}^M = -k_a^M (h^{-1})^{ab} k_b^N \gM_{NP} \partial_\alpha Y^P$ (roughly similar to the viewpoint in \cite{Eyras:1998hn}).

Now, this is particularly interesting because it seems very close to certain constructions of actions for strings and particles in doubled spacetime.
In particular, in \cite{Hull:2004in, Hull:2006va} the doubled string is reduced to the ordinary string by gauging the shift symmetry in the dual directions on which no fields depend. This gauging is implemented by introducing an auxiliary worldsheet gauge field -- let us also call it $\gV^M$ -- whose algebraic equation of motion allows one to eliminate the dual coordinates (in doing so, implementing the chirality constraint on the doubled string coordinate). More recently in \cite{Lee:2013hma, Ko:2016dxa} it was shown that this extra gauge field is also vital in order to realise the symmetries of double field theory on the worldsheet or worldline: it ensures a sort of worldvolume covariance under generalised diffeomorphisms in spacetime -- in particular, it guarantees there is a worldvolume symmetry when one has generalised Killing isometries (these ideas were extended to particles in exceptional field theory in \cite{Blair:2017gwn}). 

The gauge field $\gV^M$ is constrained to obey $\eta_{MN} \gV^M \gV^N = 0$ and $\gV^M \partial_M = 0$ (and integrating out its non-zero components the eliminates the dual coordinates from the action).
Evidently, these constraints do hold for $\gV_\alpha^M = -k_a^M (h^{-1})^{ab} k_b^N \gM_{NP} \partial_\alpha Y^P$ (assuming adapted coordinates $k_a^M \partial_M = 0$). 
This suggests that there may be a perhaps more fundamental formulation of the doubled five-brane, in which one gauges using $\gV_\alpha{}^M$, with possibly its equation of motion then ensuring the appearance of the $\hat \partial_\alpha Y^M$ we used. (Note though that in the approaches mentioned above, one uses $\gV^M$ to eliminate the components dual to the choice of section, whereas we maintained that our generalised Killing vectors could correspond to isometries beyond those mandated by the section condition. However, this is probably not difficult to reconcile.)

This suggests that one should \emph{always} view brane actions in double field theory as a sort of gauged sigma model, where one gauges away the isometry directions corresponding to the dual coordinates.
Indeed, any vector of the form $k^M = \eta^{MN} \partial_N \mathcal{O} (Y)$, where $\mathcal{O}(Y)$ is any function obeying section condition, provides a generalised Killing vector as its generalised Lie derivative is zero automatically. Then, when there is an isometry present in the physical spacetime, there is an ambiguity in the choice of whether one gauges the physical direction or its dual. This then underlies how we can obtain the KKM and the $5_2^2$ actions, as well as that of the NS5 with transverse isometries: we simply take advantage of this ambiguity, which corresponds directly to T-duality (as pointed out in for instance \cite{Berman:2014jsa} at the level of the supergravity solutions), and allows one to take some of the generalised Killing vectors to correspond to isometry directions in the physical spacetime.

A possibly related observation is to note that a generalised Killing vector $k^M = ( k, \tilde k)$ acts on the NSNS fields as
$\delta_k g = L_k g$, $\delta_k B = L_k B + d \tilde k$
where $L_k$ is the ordinary Lie derivative with respect to the vector part of $k^M$, and the one-form part $\tilde k$ generates a B-field gauge transformation. 
In the NS5 frame, we took the $d$ Killing vectors to lie solely in the dual directions, thus $k_a^M = ( 0 , \tilde k_a)$, and demanding that they are generalised Killing amounts to
\be
d \tilde k_a = 0\,.
\ee
Clearly, one can always locally introduce such $\tilde k_a$. 
If we study the NS5 brane in a background with $d$ transverse directions compactified, then one can take instead $\tilde k_a = dY^a$, where $Y^a$ are the compact coordinates on the transverse circles, which are closed but not globally exact. It is then this case that one can T-dualise in the usual sense. This suggests some interesting relationship with the (somewhat mysterious) global properties of the doubled space.

As we mentioned earlier on in the paper, ${\rm O}(d,d)$ covariant Wess-Zumino terms for solitonic branes after dimensional reduction have been provided in \cite{Bergshoeff:2011zk}, and it would be interesting to further understand how our ${\rm O}(10,10)$ expression reduced to these cases. It would also be interesting, and connected to the discussion above, to understand whether one must necessarily use the projected coordinate derivatives $\hat \partial_\alpha Y^M$ as we have done, or whether there is a more elegant formulation of the WZ term without this explicit projection.

As another comment, one should notice that the precise potential of (the dual) split ${\rm O}(d,d)$ DFT excited by the brane depends on how it is embedded into the full $D+(d+d)$ dimensional space. First sign of that is that the Bianchi identities of \cite{Bakhmatov:2016kfn} are only reproducible from the potential $D_{6,MNKL}$ in ${\rm O}(4,4)$, For other groups, which correspond to $d\neq 4$ one cannot embed the brane such as to have all fluxes internal and hence one must consider different Bianchi identities of DFT. Explicit derivation of this procedure is reserved for future work.

It would be interesting to construct the ${\rm O}(d,d)$ covariant actions for the other type II NS sector five-branes: the type IIA NS5 brane and its T-duals, presumably by starting with the PST form of the action for the IIA NS5. As double field theory can easily be extended to describe the heterotic supergravities \cite{Siegel:1993xq}, we could also study the heterotic 5-branes (the actions for the exotic versions of which were also obtained by duality in \cite{Kimura:2014upa}.

Whether the approach using generalised Killing vectors lifts to branes in exceptional field theory is a very interesting question, together with the question on applicability of the approach to the D-branes sector of Type II theory. The common subtlety of these problems is that one has to consider branes of different dimensionality which have to descend from a single action. 
In the doubled formalism, all D$p$ branes are unified into a single brane spanning half the doubled space, which appears in spacetime as D$p$ depending on how the doubled D-brane intersects the latter \cite{Hull:2004in}.
This might correspond to choosing different ways of gauging the doubled coordinates $Y^M(\sigma)$ (corresponding to the choices of the $k_a^M$), followed by a sort of partial static gauge identification so that different numbers of the coordinates actually lie in the physical and dual directions, allowing one to obtain all D$p$ branes. 

Of great interest are phenomenological applications of the obtained effective action and in particular the source-corrected Bianchi identities. One would like to consider such problems as DFT compactifications and probably involve the obtained effective action in the microstate counting of black holes, where exotic branes may well be relevant \cite{deBoer:2012ma}.

\section*{Acknowledgements}

\noindent 
CB is supported by an FWO-Vlaanderen postdoctoral fellowship,
and also by the Belgian Federal Science Policy Office through the Interuniversity Attraction Pole P7/37 ``Fundamental Interactions'', and by the FWO-Vlaanderen through the project G.0207.14N, and by the Vrije Universiteit Brussel through the Strategic Research Program ``High-Energy Physics''. 
The work of ETM is supported by the Alexander von Humboldt Foundation and in part by the Russian Government programme of competitive growth of Kazan Federal University and by DAAD PPP grant 57316852 (XSUGRA).
CB would like to thank AEI Potsdam for hospitability during the completion of this research. In return ETM thanks the Vrije Universiteit Brussel for the same. Authors would like to thank Axel Kleinschmidt, David Andriot and Alex Arvanitakis for useful discussions and comments. 

\appendix

\appendix

\section{Notations and conventions}

\subsection{Notations}

We adopt the following conventions for labelling of indices
\begin{equation}
\begin{aligned}
&\a, \b, \g, \dots =1,\dots,6&& \mbox{worldvolume} \\
&i,j,k,l,\dots =1,\dots, d && \mbox{internal space curved} \\
&\ba,\bb,\bc,\bd,\dots =1,\dots, d  && \mbox{internal space flat} \\
&\mu,\nu,\r,\s, \dots =1,\dots, 10-d  && \mbox{external space  curved}\\
&\hat{\mu}, \hat{\nu}, \hat{\r}, \hat{\s},\dots =1,\dots, 10 && \mbox{10D spacetime  curved}\\
&M,N, K, L, \dots =1,\dots, 2d  && \mbox{DFT curved, ${\rm O}(d,d)$-covariant}\\
&A,B,C,D,\dots =1,\dots, 2d && \mbox{DFT flat, {\rm O}$(d)\times${\rm O}$(d)$-covariant}\\
&a,b,c,d,\dots =1,\dots, d && \mbox{indices labelling Killing vectors}
\end{aligned}
\end{equation}

For completeness we list here notations for the objects, which appear in the paper
\begin{equation}
\begin{aligned}
DY^M&=dY^M+A_\mu^M dX^\mu, && \\
k_a^M&=(k_a^m, \tk_{am})            && \mbox{Killing vector in the generalised space}
\end{aligned}
\end{equation}

\subsection{RR field conventions}
\label{RRconv}

Here we recall the conventions of \cite{Hohm:2011dv} (related to those of \cite{Chatzistavrakidis:2013jqa} by changing the sign of $B_2$).

In IIB, we have a set of even p-forms, $C \equiv C_0 + C_2 + C_4 + \dots$, and in IIA we have odd $p$-forms $C \equiv C_1 + C_3 + \dots$. (The duals can be treated democratically in this framework. We omit them for now.) The field strengths $F$ are defined as
\begin{equation}
\begin{array}{ccl}
F_1  & = & d C_0 \\
F_3 & = & dC_2 + C_0 H_3 \\
F_5 & = & dC_4 + C_2 \wedge H_3 \\
 & = & d C_4^\prime + \frac{1}{2} C_2 \wedge H_3 - \frac{1}{2} B_2 \wedge dC_3 
\end{array} 
\begin{array}{ccl}
F_2 & = & dC_1 \\
F_4 & = &d C_3 + H_3 \wedge C_1  \\
 & & \\
 & & \\
\end{array}
\label{RRF}
\end{equation}
where
\begin{equation}
C_4^\prime = C_4 + \frac{1}{2} C_2 \wedge B_2 \,,
\end{equation}
is an alternative choice for the four-form potential used in some papers. 
The total gauge transformations of these fields are:
\begin{equation}
\begin{array}{ccl}
\delta C_0 & = & 0 \\
\delta C_2 & = & d \lambda_1 \\
\delta C_4 & = & d\lambda_3 - d\lambda_1 \wedge B_2 \\
\delta C_4^\prime &=& d\lambda_3 - \frac{1}{2} d\lambda_1 \wedge B_2 + \frac{1}{2} d \Lambda_1 \wedge C_2 \\
\delta B_2 & =& d \Lambda_1
\end{array}
\begin{array}{ccl}
\delta C_1 & = & d \lambda_0 \\
\delta C_3 & = & d\lambda_2 - B_2 \wedge d \lambda_0\\
 & & \\
 & & \\
 & & \\
\end{array}
\end{equation}
So note that the IIB fields $(C_0,C_2,C_4)$ are invariant under B-field gauge transformations (these are denoted $\hat A_p$ in \cite{Hohm:2011dv}) while $(C_0, C_2, C_4^\prime)$ are not (these are denoted $A_p$ in \cite{Hohm:2011dv}). 

Following \cite{Hohm:2011dv}, we define an alternative set of p-form potentials $\cC_p$ (in \cite{Hohm:2011dv} these are what they denote by $C_p$) by
\begin{equation}
\cC = e^{B_2} \wedge C\,,
\label{calC}
\end{equation}
so that
\begin{equation}
\begin{array}{ccl}
\cC_0 & =& C_0 \,,\\
\cC_2 & =& C_2+ B_2 C_0 \,,\\
\cC_4 & =& C_4 + B_2 \wedge C_2 + \frac{1}{2}B_2 \wedge B_2 C_0 \,,\\
\end{array} \qquad
\begin{array}{ccl}
\cC_1 & =& C_1 \,,\\
\cC_3 & = & C_3 + B_2 \wedge C_1 \,.\\
 & & \\
 \end{array}
\end{equation}
The gauge transformations are
\begin{equation}
\begin{array}{ccl}
\delta \cC_0 & = & 0\\
\delta \cC_2 & = & d \lambda_1 + d \Lambda_1 \cC_0 \\
\delta \cC_4 & = & d \lambda_3 + d\Lambda_1 \wedge \cC_2 
\end{array}\qquad
\begin{array}{ccl}
\delta C_1 & = & d \lambda_0 \\
\delta C_3 & = & d\lambda_2 + d \Lambda_1 \wedge \cC_1\\
 & & \\
\end{array}
\end{equation}
so these are not invariant under B-field transformations, but however transform in a consistent manner under them.
In \cite{Hohm:2011dv} they show that it is in fact the potentials $\cC_p$ which are encoded as ${\rm O}(d,d)$ spinors. 

\subsection{NS5 worldvolume conventions} 

To describe the NS5 brane, we also need to introduce some dual potentials in IIB. We have
\be
\begin{split} 
F_7 & = d C_6 + C_4 \wedge H_3 \\
\delta C_6 & = d \lambda_5 - d\lambda_3 \wedge B_2 + \frac{1}{2} d \lambda_1 B_2 \wedge B_2 
\end{split} 
\ee
while the field strength of 6-form dual to the B-field is defined via
\be
dH_7 = - F_3 \wedge F_5 + F_1 \wedge F_7 \,.
\ee
We solve this Bianchi identity via (after \cite{deBoer:2012ma} with respect to whom we have $B_2 \rightarrow - B_2$):
\be
H_7 = dB_6 - \frac{1}{2} \left( F_1 \wedge C_6 - F_3 \wedge C_4 + F_5 \wedge C_2 - F_7 \wedge C_0 \right) \,.
\ee
A short calculation shows that the gauge transformations of $B_6$ can be taken to be
\begin{equation}
\begin{aligned}
\delta B_6 =&\ d \Lambda_5 + \frac{1}{2} \Big(
- F_1 \wedge ( \lambda_5 - B_2 \wedge \lambda_3 + \frac{1}{2} B_2 \wedge B_2 \wedge \lambda_1 ) \\
&+ F_3 \wedge (\lambda_3 - B_2 \wedge \lambda_1) - F_5 \wedge \lambda_1 
\Big) \,.
\end{aligned}
\end{equation}
To construct a gauge invariant WZ term, we introduce worldvolume gauge potentials $c_1,c_3,c_5$ and their field strengths:
\be
\begin{split} 
\mG_0 & = C_0 \,,\\
\mG_2 & = d c_1 + C_2 \,, \\
\mG_4 & = d c_3 + H_3 \wedge c_1 + C_4 \,, \\
\mG_6 & = d c_5 + H_3 \wedge c_3 + C_6 \,.
\end{split} 
\ee
The potentials transform as 
\be
\begin{split}
\delta c_1 & = - \lambda_1 \\
\delta c_3 & = - \lambda_3 + \lambda_1 \wedge B_2 \\
\delta c_5 & = - \lambda_5 + \lambda_3 \wedge B_2 - \frac{1}{2} \lambda_1 \wedge B_2 \wedge B_2 \,.
\end{split}
\ee
The Wess-Zumino term is (this is the expression in \cite{deBoer:2012ma}. Two of the terms appear to be missing from the expression in \cite{Bergshoeff:2011zk}):
\be
\mathcal{L}_{WZ} = B_6 + \frac{1}{2} \left( \mG_6 C_0 - \mG_4 \wedge C_2 + \mG_2 \wedge C_4 - \mG_0 C_6 \right) \,.
\label{WZ}
\ee
Under gauge transformations, we have
\begin{equation}
\begin{aligned}
&\delta \mathcal{L}_{WZ} =\\
&\ d \left[ \Lambda_5 + \frac{1}{2} \left( - \mG_4 \wedge \lambda_1  + \mG_2 ( \lambda_3 - B_2 \wedge \lambda_1 ) 
- \mG_0 ( \lambda_5 - B_2 \wedge \lambda_3 - \frac{1}{2} B_2 \wedge B_2 \wedge \lambda_1 )
\right) \right] \,.
\end{aligned}
\end{equation}
The expression \eqref{WZ} can also be written as 
\be
\mathcal{L}_{WZ} = B_6 + \frac{1}{2} \left(
( d c_5 + H_3 \wedge c_3 ) C_0
- ( dc_3 + H_3 c_1 ) \wedge C_2 
+ d c_1 \wedge C_4 
\right) \,.
\ee
If we let 
\be
\begin{split}
\tilde c_5 & = c_5 + B_2 \wedge c_3 + \frac{1}{2} B_2 \wedge B_2 \wedge c_1 \,,\\
\tilde c_3 & = c_3 + B_2 \wedge c_1\,,\\
\tilde c_1 & = c_1 \,,
\end{split}
\ee
transforming as $\delta \tilde c_p = - \lambda_p$,
then we further see that
\be
\mathcal{L}_{WZ} = B_6 + \frac{1}{2} \left( d \tilde c_5 \cC_0 - d \tilde c_3 \wedge \cC_2 + d \tilde c_1 \wedge \cC_4 \right) \,,
\ee
with the calligraphic RR forms as defined in \eqref{calC}. 
Alternatively, one can define
\be
\tilde \mG_p = d \tilde c_{p-1} + \cC_p 
\ee
then the WZ term is also expressible as:
\be
L_{WZ} = B_6 + \frac{1}{2} \left(
\tilde \mG_6 \cC_0 
-\tilde \mG_4 \cC_2
+\tilde \mG_2 \cC_4
-\tilde \mG_0 \cC_6 
\right) \,.
\ee

\subsection{T-duality of RR fields} 
\label{RRT}

The following $\mathrm{Pin}(d,d)$ transformation:
\be
S_i \equiv \psi^i + \psi_i 
\ee
(with $S_i^{-1} = S_i$) induces the following ${\rm O}(d,d)$ tranformation \cite{Hohm:2011dv}
\be
h_i = \begin{pmatrix} 
 - 1 + e_i & e_i \\
e_i & -1 + e_i 
\end{pmatrix} 
\quad,\quad
(e_i)_{jk} = \delta_{ij} \delta_{ik} \,.
\ee
Acting on the generalised coordinates $Y^M$, this interchanges $Y^i$ with $\tilde Y_i$ and sends $Y^M \rightarrow - Y^M$ for $Y^M \neq Y^i, Y_i$. 
Thus it generates a Buscher transformation in the direction $i$ accompanied by a reflection in all other directions. 

We find that acting on a spinor $\cC$ we have, taking $i=1$ for definiteness,
\be
\begin{split} 
 (S_1 \cC)_{ \mu_1 \dots \mu_p }& = \cC_{1 \mu_1 \dots \mu_p }\\
 (S_1 \cC)_{ 1 \mu_1 \dots \mu_p }& = \cC_{ \mu_1 \dots \mu_p }\,,
\end{split} 
\ee
where $\mu_i$ indices run over the directions excluding $i=1$.
Taking into account the reflections, this implements the Buscher rules as:
\be
\begin{split} 
 \tilde \cC_{ \mu_1 \dots \mu_p }& = (-1)^p\cC_{1 \mu_1 \dots \mu_p }\\
 \tilde \cC_{ 1 \mu_1 \dots \mu_p }& = (-1)^p \cC_{ \mu_1 \dots \mu_p }\,,
\end{split} 
\ee
In particular, IIA and IIB are related very simply via:
\be
\begin{split} 
\cC_1 & = \cC_0 \,,\\
\cC_\mu & = \cC_{\mu 1}\,, \\
\cC_{\mu \nu 1} & = \cC_{\mu \nu}\,, \\
\cC_{\mu\nu\rho} & = \cC_{\mu\nu\rho 1}\,, \\
\end{split} 
\ee
and so on for the higher rank potentials. From this one can find, for instance, that the components of the ten-dimensional IIA RR fields are related to those of the IIB fields by:
\be
\begin{split}
 \hat C_1 & = C_0 \\
 \hat C_\mu & = \hat C_{\mu 1 } + \hat B_{\mu 1} C_0 \\
 \hat C_{\mu\nu 1} & = \hat C_{\mu\nu} + 2 A_{[\mu}{}^1 \hat C_{\nu]1} \\
\hat C_{\mu\nu\rho} & = \hat C_{\mu\nu\rho 1} + 3 \hat B_{[\mu |1|} \hat C_{\nu \rho]} 
 - 6 A_{[\mu}{}^1 \hat B_{\nu|1|} \hat C_{\rho]1}\,.
\end{split} 
\label{RRTduality} 
\ee
This agrees with the result of reducing the field strengths of \eqref{RRF} to 9 dimensions and matching the resulting (appropriately redefined) components.

The result of doing two T-dualities follows either by inverting \eqref{RRTduality} to find the rules for the IIB fields, or by applying $(\psi^1 + \psi_1) (\psi^2 + \psi_2)$ to the spinor and then acting with a reflection in both the $1$ and $2$ directions. One easily obtains the T-duality transformations:
\be
\begin{split}
\tilde \cC_0  & = - \cC_{12}\\
\tilde \cC_{12} & = \cC_0 \\
\tilde \cC_{\mu 1} & = - \cC_{\mu 2} \\
\tilde \cC_{\mu 2} & = \cC_{\mu 1} \\
\tilde \cC_{\mu \nu} & = \cC_{\mu \nu 12} \\
\tilde \cC_{\mu\nu 12} & = \cC_{\mu \nu} \,,
\end{split}
\label{2Ts}
\ee
and so on.

\bibliographystyle{JHEP}
\bibliography{bib}
\end{document}